\newif\ifTWC
\TWCfalse

\ifTWC
    \documentclass[12pt, draftclsnofoot, onecolumn]{IEEEtran}
\else
    \documentclass[journal]{IEEEtran}
\fi

\usepackage[T1]{fontenc}     
\usepackage{newtxtext}       
\usepackage{newtxmath}      
\usepackage{courier}        
\usepackage[mathscr]{eucal}  
\usepackage{CJK} 

\usepackage{capt-of}
\usepackage{amsmath,amsfonts}
\usepackage{algorithmic}
\usepackage{algorithm}
\usepackage{array}
\usepackage{makecell}
\usepackage[caption=false,font=normalsize,labelfont=sf,textfont=sf]{subfig}
\usepackage{textcomp}
\usepackage{stfloats}
\usepackage{url}
\usepackage{verbatim}
\usepackage{float}
\usepackage{graphicx}
\usepackage{cite}
\usepackage{booktabs}
\usepackage{color}
\usepackage{balance}  
\usepackage[table,xcdraw]{xcolor}
\usepackage{multirow} 
\usepackage{booktabs}   
\usepackage{xcolor} 
\definecolor{mycolor}{RGB}{0, 0, 255} 

\newcommand{\bluesection}[1]{%
    \section{\textcolor{blue}{#1}}%
}

\begin{document}

\title{\LARGE Design and Analysis of Phase Conjugation-Based Self-Alignment Beamforming for RIS-Assisted Terahertz SWIPT}

\author{
    \normalsize Jiayuan Wei, Qingwei Jiang, Wen Fang, Mingqing Liu, Qingwen Liu,~\IEEEmembership{Senior Member,~IEEE},\\ Wen Chen,~\IEEEmembership{Senior Member,~IEEE}, and Qingqing Wu,~\IEEEmembership{Senior Member,~IEEE}

    \thanks{
        J. Wei, Q. Jiang, and Q. Liu
        are with the College of Computer Science and Technology, Tongji University, Shanghai 201804, China
        (e-mail: jywei@tongji.edu.cn, jiangqw@tongji.edu.cn, qliu@tongji.edu.cn).
    }
    
    \thanks{
        W. Fang and M. Liu are with the College of Electronic and Information Engineering, Tongji University, Shanghai 201804, China (e-mail: wen.fang@tongji.edu.cn, clare@tongji.edu.cn).
    }
    \thanks{W. Chen and Q. Wu are with the Department of Electronic Engineering, Shanghai Jiao Tong University, Shanghai 200240, China (email: wenchen@sjtu.edu.cn, qingqingwu@sjtu.edu.cn).
    }
       
}

\maketitle





\begin{abstract}

\textcolor{blue}{Terahertz (THz) simultaneous wireless information and power transfer (SWIPT) is a promising technology for enabling ultra-high-rate and low-latency communications in massive battery-free Internet of Things (IoT) deployments for 6G networks. However, conventional THz systems rely on narrow directional beams that necessitate precise alignment, typically achieved through high-overhead beam scanning procedures, which fundamentally at odds with the energy constraints of battery-free IoT devices. In this paper, we propose a novel self-alignment architecture for THz SWIPT leveraging a reconfigurable intelligent surface (RIS) to eliminate complex beam scanning. By integrating phase conjugate circuits at both the base station and user equipment, the RIS facilitates a resonance-based bidirectional retro-reflection mechanism, enabling the system to autonomously converge to an aligned state without manual intervention. We develop an analytical channel transfer model and a power cycle model to characterize the resonance-assisted beam alignment process and power transfer efficiency. Simulation results demonstrate that the RIS-enabled system achieves effective spatial power concentration with significant sidelobe suppression, leading to a communication capacity of 127.84 Gbit/s and a received power of 13.62 mW over a 2.2-meter link.}
\end{abstract}

\begin{IEEEkeywords}  
Terahertz, reconfigurable intelligent surface, simultaneous wireless information power and transfer, beam alignment.
\end{IEEEkeywords}

\section{INTRODUCTION}

\color{blue}\IEEEPARstart{T}{he} simultaneous wireless information and power transfer (SWIPT) has become a key enabling technology for sustainable and large-scale connectivity in future wireless networks\cite{r1,r2}, particularly with the rapid growth of energy-autonomous applications in the Internet of Things (IoT), wearable electronics, and cyber-physical systems. However, the development of conventional SWIPT systems operating in lower frequency bands are increasingly constrained by limited available spectrum. This limitation has motivated a transition to the high frequency, e.g. terahertz (THz) band (0.1-10 THz), whose ultra-wide bandwidth supports high-speed data transmission and efficient wireless power delivery \cite{r3}. THz-SWIPT thus emerges as a pivotal technology for realizing the 6G vision of pervasive intelligent connectivity, demonstrating substantial potential for deployment in integrated sensing and communication \cite{r4}, vehicle-to-vehicle communication \cite{r5}, and nanoscale IoT devices \cite{r6}.\color{black}

\textcolor{blue}{Terahertz technology offers unique advantages that make it particularly attractive for next-generation wireless systems. However, one of the key challenges in establishing THz links is beam misalignment \cite{r7,r8}. THz signals suffer from high path loss and are highly susceptible to molecular absorption, which necessitates the use of narrow beams to concentrate limited energy toward intended receivers while achieving beamforming gain \cite{r9}. Unfortunately, such highly directional beams require beam scanning procedures between transceivers for alignment. This process typically involves an exhaustive search across the angular domain to identify the optimal beam pair that maximizes signal strength and quality, which incurs unaffordable training overhead. Therefore, efficient beam alignment is crucial for practical THz wireless communications.}

\textcolor{blue}{Reconfigurable intelligent surfaces (RIS) have demonstrated remarkable advantages in beam alignment due to their programmable phase-shifting capability. By dynamically configuring the phase shifts of a large number of passive reflecting elements, RIS can constructively combine the incident signals toward the intended receiver, thereby enabling highly directional beamforming, while also improving spectral efficiency, and concentrating signal energy in desired directions \cite{r11,r28}. In RIS-assisted THz beam alignment, commonly adopted methods include exhaustive search, hierarchical search, multi-beam search and others, as presented in Table \ref{research}, however, still operate within the conventional paradigm of active beam scanning. For example, exhaustive search is the most straightforward and accurate beam alignment strategy, in which the UE needs to scan the entire angle space for evey combination of transmitting and reflecting beam patterns \cite{r46},\cite{r47}, results in exceedingly high training complexity and time overhead \cite{r10}. To reduce training complexity, hierarchical search is explored. For instance, \cite{r12} proposed a full-coverage hierarchical beam search scheme that probes multiple angles simultaneously at each layer and identifies the optimal beam direction by synthesizing detection results across all layers. Another study \cite{r13} designed a two-phase hierarchical beam training protocol assisted by RIS, which optimizes RIS phase configurations and alternates between coarse-grained and fine-grained scanning to achieve beam alignment with low system overhead. Nevertheless, the performance of hierarchical search heavily depends on codebook design, and spatial resolution increases with the number of layers.  Multi-beam search has attracted attention due to its high search efficiency. For example, P. Wang et al. developed a multi-beam training framework \cite{r14}, where the transmitter and RIS collaboratively generate multiple narrow beams for concurrent angular scanning, followed by an intersection strategy to identify the optimal beam pair. Similarly, W. Mei et al. proposed an RIS-based multi-path beam routing scheme that employs beam splitting and merging in \cite{r15}. In this method, the BS transmits signals to the RIS via active beams, and the RIS reflects and combines the signals via passive beamforming, ultimately achieving coherent beam reception and alignment at the UE. This approach rapidly narrows down the search space for the optimal beam by simultaneously probing multiple directions. Furthermore, other methods also have been studied in \cite{r48}, a distributed beam training method that combines offline and online phases by leveraging coordinated training between the BS and the RIS controller.}

\textcolor{blue}{The aforementioned methods, while reducing overhead to varying degrees, fundamentally rely on structured scanning procedures which remain a source of latency and complexity. This inherent limitation calls for a paradigm shift from \textit{active scanning} to \textit{passive self-alignment}. Recently, the resonance mechanism exhibiting autonomous beam alignment characteristics has been proposed and demonstrated in extensive applications in both optical and radio frequency systems \cite{r19, r20, r21}. By employing retro-reflectors on both the BS and the UE, the beam can automatically align the UE, realizing self-alignment. Xiong et al. \cite{r22} proposed an adaptive communication system based on the resonance mechanism to achieve high-power laser mobile communications, enabling self-alignment between the BS and mobile devices without requiring active motion tracking. Subsequently, Fang et al. \cite{r23} presented RIS-assisted SWIPT performance based on the resonance mechanism, with systematic analysis demonstrating the received power and energy efficiency metrics in non-line-of-sight self-alignment configurations.}

\begin{table*}[htbp]\color{blue}
\centering
\caption{Comparison of beam alignment techniques}
\renewcommand{\arraystretch}{1.2} 
\begin{tabular}{c|c|c|c}
\hline
\textbf{Techniques} & \textbf{Ref.}  & \textbf{Features and analysis} &\textbf{self-alignment} \\
\hline
\multirow{2}{*}{Exhaustive search} & \cite{r46} &   \multirow{2}{*}{\parbox{10cm}{It can identify the optimal beam direction, but incurs substantial training overhead. }}  & \multirow{2}{*}{No} \\
\cline{2-2} 
 & \cite{r47}  &  \\
\hline
\multirow{2}{*}{Hierarchical search} & \cite{r12}  &\multirow{2}{*}{\parbox{10cm}{It can achieve high search efficiency, but the performance heavily relies on the design of the training codebook and the spatial resolution increases with the number of layers. }}  &\multirow{2}{*}{ No}   \\
\cline{2-2} 
 & \cite{r13}  &   \\
\hline
\multirow{2}{*}{Multi-beam search} & \cite{r14}   & \multirow{2}{*}{\parbox{10cm}{It can accelerate the beam alignment process and reducing training time, but the potentially scan codebook size at the RIS increase the training complexity. }}  & \multirow{2}{*}{No}  \\
\cline{2-2} 
 & \cite{r15}  &   \\
\hline
Other & \cite{r48}  &  \multirow{1}{*}{\parbox{10cm}{It provides a distributed beam training technique instead of exhaustive search. }} &No \\
\hline
\textbf{self-alignment} & \textit{this work}  &  \multirow{1}{*}{\parbox{10cm}{It can achieve autonomous beam alignment without beam scanning procedures.}} & \textbf{Yes} \\
\cline{2-2} 
\hline
\end{tabular}
\label{research}
\end{table*}

\textcolor{blue}{Thus, in this paper, we propose a THz self-alignment SWIPT system designed to eliminate the need for complex beam scanning procedures, thereby supporting autonomous beam alignment toward UE. Furthermore, when the UE moves within the field of view (FoV), the BS can autonomously track its mobility and rapidly re-establish an aligned link. Unlike prior studies \cite{r24,r25}, which explored self-conjugating beamforming, where signals are retro-reflected by a conjugated RIS back to the BS for demodulation before being scattered to the UE, the proposed method establishes a end-to-end self-alignment link thorough bidirectional retro-reflection. This enables coherent superposition of signals in the wave domain through mutual retro-reflection. From a wave physics perspective, the superimposed signal remains stable and invariant after multiple reflections, thereby naturally aligning directly with the UE and establishing a beam-aligned communication link without the need for beam scanning.}

\begin{figure*}[h]\color{blue}
\centering
\includegraphics[height = 2.6in,width=4.8in]{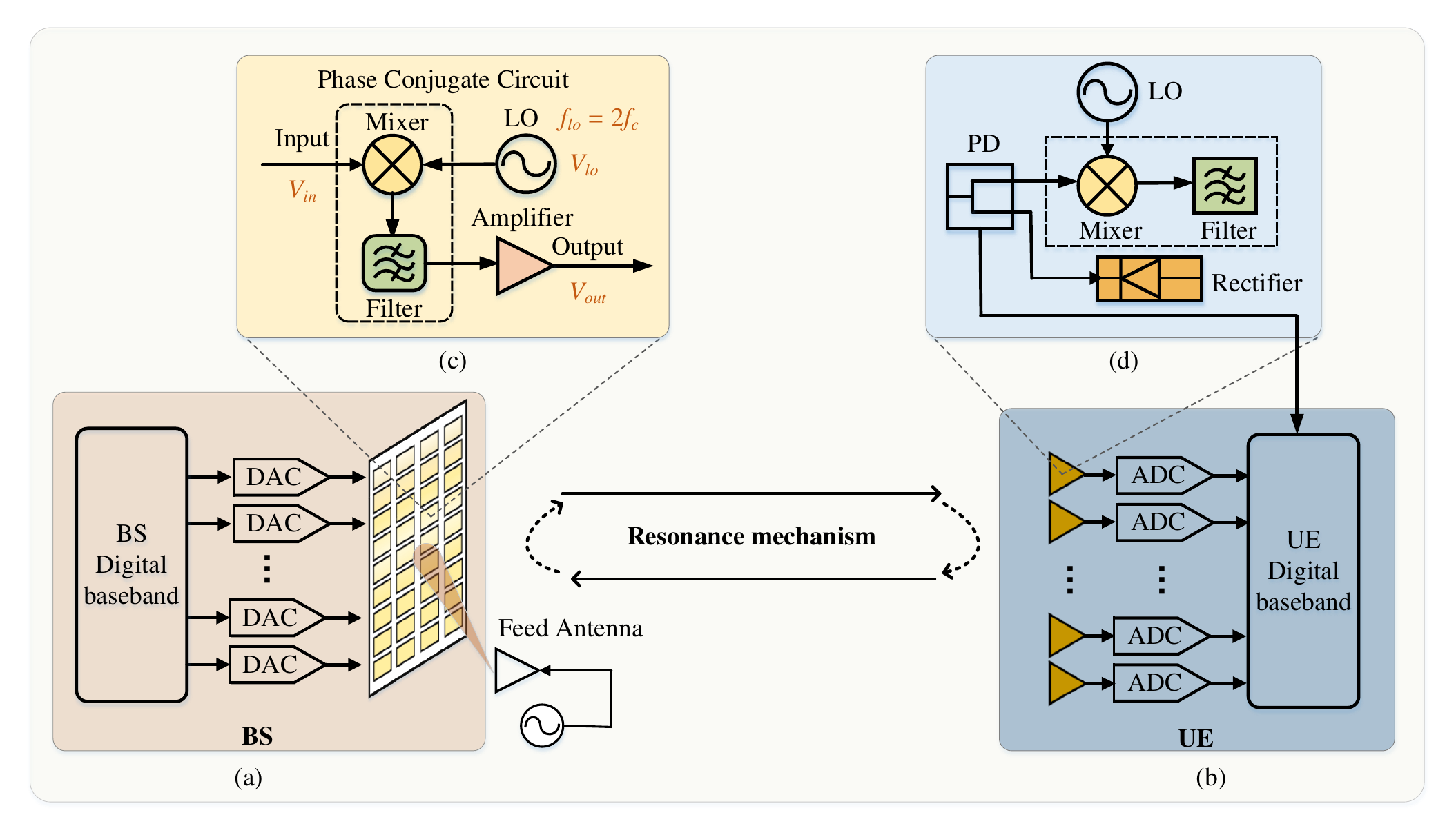}
\caption{An exemplary RIS-assisted THz self-alignment SWIPT system.}
\label{fig: architerture}
\end{figure*}

\textcolor{blue}{The main contributions are summarized as follows.}
\begin{itemize}

\item[$\bullet$]\textcolor{blue}{We propose a resonant mechanism-based RIS-assisted THz SWIPT system capable of autonomous beam self-alignment for UE moving within the FoV. In this architecture, the RIS serves as the transmitter, integrated with a phase conjugate circuit and a power amplifier, while an antenna array equipped with a phase conjugate circuit functions as the receiver. This configuration enables bidirectional retro-reflection, wherein the signal is retro-reflected from the UE to the BS and back to the UE iteratively. Within each iteration, the signal power is amplified to compensate for the path loss. Through multiple iterations, the system autonomously establishes a stable communication link, achieving beam self-alignment without relying on complex and explicit beam scanning procedures.}

\item[$\bullet$] \textcolor{blue}{We developed a THz channel transmission model and a power cycle model to characterize the behavior of the proposed system. Based on the Friis free-space transmission equation, the THz signal transmission model is established to quantify the received power and transfer efficiency. And the power cycle model captures the dynamic power fluctuation between the BS and the UE throughout the retro-reflective iterations. We further analysed steady-state conditions under which the power gain by amplified exactly compensates the path loss, so that the power distribution converges and no longer varies, indicating successful establishment of a stable power cycle.}

\item[$\bullet$] \textcolor{blue}{Numerical simulations validate the proposed system successfully accomplishes beam self-alignment and focusing. Throughout resonance mechanism establishment, significant sidelobe suppression is observed, resulting in an energy-concentrated communication channel that optimally directs transmitted power toward the intended UE. The simulation traces the spatial evolution of the signal from initial divergence to final focused state, including detailed power distribution and SWIPT performance metrics. Simulation results show the proposed system demonstrates a maximum effective transmission range of 2.2 meters while maintaining a power transfer efficiency of 68.1\%, with a delivered power of 13.62 mW.} 
        
\end{itemize}

\textcolor{blue}{The paper is structured as follows. Section II details the system architecture, operational principles of the phase conjugate circuit, and the self-alignment mechanism. Section III describes the RIS structural features and THz channel characteristics, followed by derivation of the power steady-state cycle model. A comprehensive analysis of the SWIPT performance is provided in Section IV. Experimental results and discussions are presented in Section V. Finally, Section VI concludes the study and outlines potential future research directions.}

\section{SYSTEM OVERVIEW}

\subsection{System Architecture}

\textcolor{blue}{
The resonance mechanism-based RIS-assisted THz SWIPT system is shown in Fig.  \ref{fig: architerture}, Specifically, Fig.  \ref{fig: architerture}(a) depicts the BS equipped with a RIS-assisted transmitter, which comprises a feed antenna, a digital baseband module, and a digital-to-analog converter (DAC). The digital baseband module generates the corresponding baseband digital sequence, which are then converted by the DAC into bias voltage signals that control the phase response of the RIS. The initial signal is emitted by the feed antenna to the RIS. At the UE side, as shown in Fig. \ref{fig: architerture}(b), the receiver is configured with a dedicated antenna \cite{r26}, and mainly consists of an analog-to-digital converter (ADC) and a baseband module, responsible for signal digitization and information decoding, respectively. }

\textcolor{blue}{Furthermore, the structural composition of each unit of the BS and UE is represented in Fig. \ref{fig: architerture}(c) and Fig. \ref{fig: architerture}(d), respectively. In Fig. \ref{fig: architerture}(c), each unit cell of the RIS consists of a phase conjugate circuit and a PA. The phase conjugate circuit ensures the conjugation of the signal phase, while the power amplifier (PA) provides the necessary gain for establishing a stable power cycle for the signal. In Fig. \ref{fig: architerture}(d), the antenna unit consists of a power divider (PD), a phase conjugate circuit and a rectifier. The PD manages the power received by the UE, allocating a portion for power rectifier while simultaneously directing a fraction of the power back to the BS, and the remaining is for UE digital baseband for information demodulation. The signal is reflected from the RIS to the BS, undergoes phase conjugation and retro-reflection, and finally returns to the BS is defined as an iteration.  After several iterations, the phases of the unit elements align with each other, and the beam converges to a steady state that remains unchanged in subsequent iterations, which is referred to as the resonance mechanism.}

\subsection{Phase Conjugate}

\textcolor{blue}{The feasibility of implementing phased arrays capable of conjugating the phase of incident waves in the microwave band has been experimentally demonstrated in several earlier studies\cite{r26}. The phase conjugate circuit generates the opposite phase of the input signal. }

\begin{equation}E=A(r)e^{\pm j(kr+\omega t)},\end{equation}
\textcolor{blue}{where ${A}$ denotes the amplitude, ${r}$ is the position vector, ${k} = 2\pi/\lambda$ represents the wave number vector indicating propagation direction, $\lambda$ is the wavelength, and ${\omega}$ is the angular frequency. }

\textcolor{blue}{In \cite{r27}, a solution utilizes an antenna array combined with mixing between the incident wave (at center frequency $f_{c}$) and a locally generated sinusoidal wave at $2f_{c}$ to achieve phase conjugation. More recently, it has been shown that reconfigurable retrodirective metasurface is proposed and verified through experiments\cite{r29}. In this work, the operating principle of the phase conjugate circuit, as shown illustrated in Fig. \ref{fig: architerture}(c), follows the same physical mechanism as that introduced in \cite{r27}. we can observe that the phase conjugate circuit operates by mixing the input signal with a local oscillator (LO) signal. This process generates both sum and difference frequency components. A low-pass filter is then used to eliminate the sum-frequency component, retaining the difference-frequency signal as the output. The resulting output has the same frequency as the carrier signal but exhibits a phase conjugate to it. Specifically, let the incident signal be defined as $V_{in} = A_{c}\cos(2\pi f_{c}t+\varphi)$, and the LO signal as $V_{lo} = A_{lo}\cos(2\pi f_{lo}t)$, where $A_{c}$ and $A_{l}$ denote the signal amplitudes, $\varphi$ is the phase offset, and $f_{lo} = 2f_{c}$ is the LO frequency used in the mixer. Considering the $m$-th element at the BS, the mixing process between the incident signal and the doubled-frequency LO signal can be expressed as}

\begin{equation}\color{blue}
\begin{aligned}
V_{if} &= \frac{1}{2}A_{c}A_{lo} \left[ \cos{ \left( 2\pi(f_{lo}-f_{c}) - \varphi_{m} \right) } \right. \\
&\quad \left. + \cos{ \left( 2\pi(f_{lo}+f_{c}) + \varphi_{m} \right) } \right].
\end{aligned}
\end{equation}

\textcolor{blue}{The signal component at frequency  $f_{lo}+f_{c}$ can be readily removed using a band-pass filter. Consequently, the signal incident on the $m$-th element is given by}

\begin{equation}\color{blue}
\begin{aligned}
 V_{out} &= \frac{1}{2}A_{c}A_{}[\cos{(2\pi(f_{lo}-f_{c})t-\varphi_{m}})] \\
 & = \frac{1}{2}A_{lo}A_{c}\cos{(2\pi f_{c}t-\varphi_{m})}.
 \end{aligned}   
\end{equation}

\textcolor{blue}{It can thus be seen from the equation that the output signal shares the same frequency as the input signal but has a opposite phase. Additionally, the phase conjugate circuit introduces an amplitude attenuation factor of $A_{lo}/{2}$}.

\subsection{Self-Alignment Mechanism}
\textcolor{blue}{The signal transmission process in the system comprising the following steps: (i) A carrier signal is emitted from the feed antenna and illuminates the RIS, with the phase is set through random initialization, the signal flows into  phase conjugate circuit and is amplified by the PA then scattered to the UE. (ii) At the UE, the PD manages the power received by the UE, the majority of the received power is harvested by the power rectifier and converted into direct current (DC) for device charging, while a small portion is captured by the UE digital baseband for information reception, and the remaining small part of power is used to drive the embedded phase conjugate circuit, which reflects the signal back to the BS. (iii) Upon reception at the BS, the signal is amplified and phase-conjugated before being retro-reflected to the UE. When it returns to the UE, part of the signal is fed back again, repeating the process until a steady power cycle is established. (iv) Finally, the DAC converts the digital baseband sequence into an analog voltage signal to control each unit cell of the RIS, achieving phase modulation of the steady-state signal. The modulated signal is transmitted along the stabilized path, and the ADC at the UE reconstructs the digital signal for the information decoding.}

\begin{figure}[h]\color{blue}
    \centering
    \includegraphics[width=3in,height = 1in]{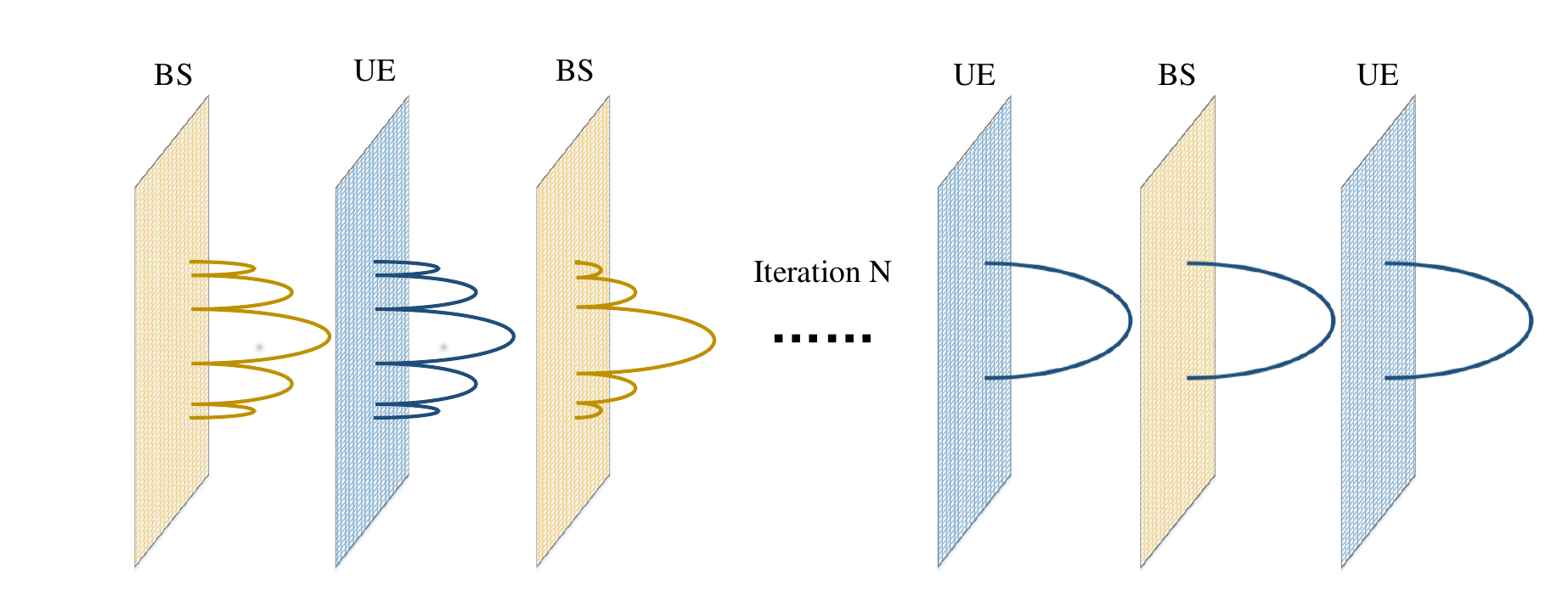}
    \caption{Iterative process of the self-reproducing mode}
    \label{fig:self-reproucing}
\end{figure}

\textcolor{blue}{During the initial iteration phase,  the transmitted signal experiences significant beam divergence and fails to align accurately with the UE, resulting in pronounced sidelobes. As the number of iterations increases, the signals gradually achieve coherent superposition between the BS and the UE. The phase distributions at both ends converge to a consistent pattern, which sharpens the main lobe toward the UE center and effectively suppresses sidelobe effects. This process markedly reduces path loss and enhances power transfer efficiency. The system is considered to have reached a steady state when the power variation of the array elements between two consecutive iterations falls below a negligible value. At this point, a stable power distribution emerges and repeats itself, a behavior defined as the self-reproducing mode. As shown in Fig. \ref{fig:self-reproucing}, the signal evolves from an initial state characterized by prominent sidelobes and scattered energy into a focused and stable beam profile that consistently reproduces itself across multiple iterations.}

\begin{figure}[h]
\centering
\includegraphics[height =1.6in,width=1.6in]{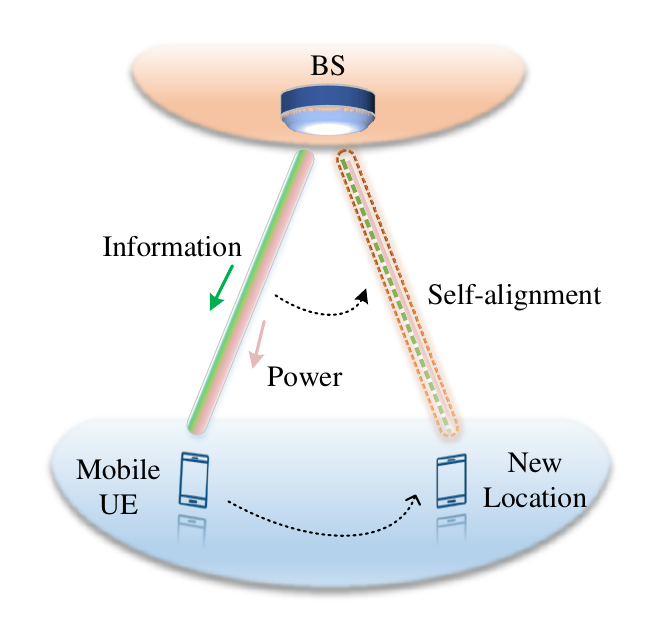}
\caption{The self-alignment link re-establishment of mobile UE.}
\label{fig: self-alignment}
\end{figure}

\textcolor{blue}{Once the self-reproducing mode is attained, the self-alignment mechanism is successfully established. This self-alignment capability emerges spontaneously through the use of physical retro-reflective structures, thereby eliminating the reliance on complex beam scanning algorithms. Moreover, even when the UE moves within the FoV, the transceiver can rapidly re-establish a self-aligned link and realign the signal direction toward the UE. As depicted in Fig. \ref{fig: self-alignment}, when the receiver is displaced, the inherent retro-reflective mechanism allows the transmitter to directionally track the receiver and autonomously reconverge to a new steady state, thereby re-establishing self-alignment without external intervention.}

\section{TRANSFER MODEL}

\textcolor{blue}{This section details the design of the RIS-assisted transmitter, analyzes the THz channel gain, and develops the THz transfer model and resonance mechanism for stable power cycle. The energy harvesting and communication channel models are also analyzed.}

\subsection{Fundamentals of RIS-assisted Transmitter}

\begin{figure}[h]
\centering
\includegraphics[width=3.2in,height = 1.8in]{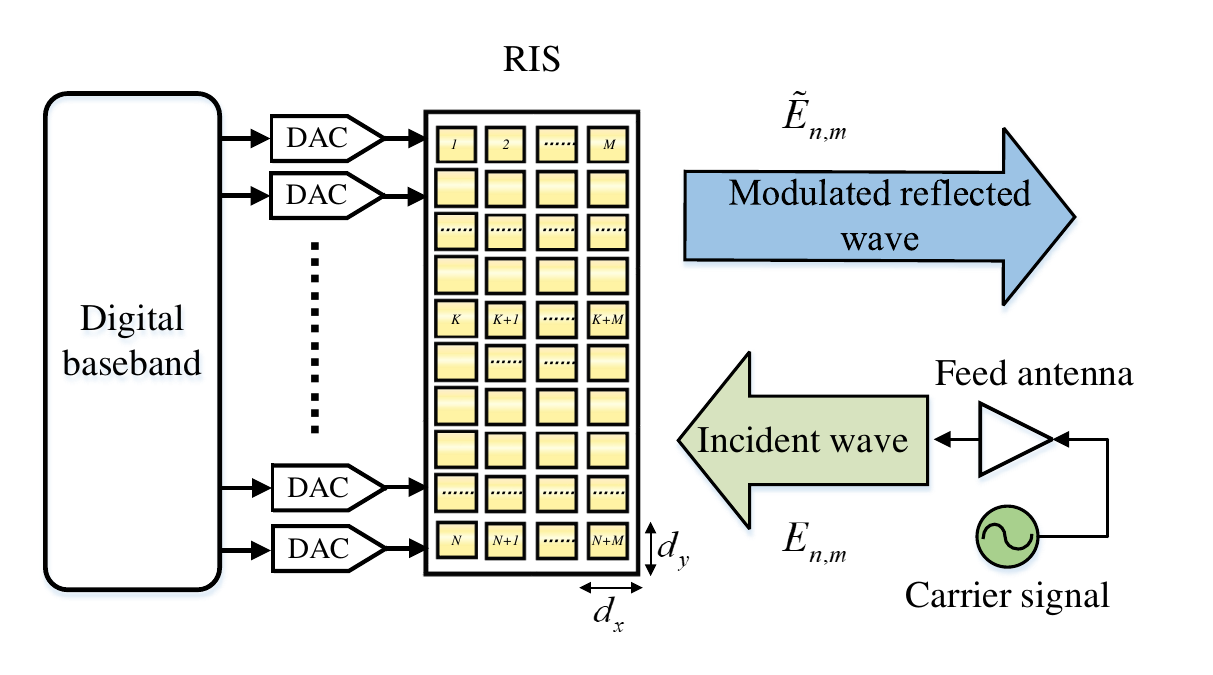}
\caption{A RIS-assisted transmitter prototype.}
\label{fig: RIS-Intro}
\end{figure}

\textcolor{blue}{ We consider a RIS-assisted structure arranged in a regular rectangular grid, as shown in Fig. \ref{fig: RIS-Intro}, the baseband module is directly connected to each unit cell of RIS.} Let $N$ and $M$ represent the number of RIS unit cells along the vertical and horizontal dimensions, respectively. The unit cell $U_{n,m}$ is located in the $n$-th row and $m$-th column of the array, with each unit defined by physical dimensions $d_x$ and $d_y$. Taking the unit cell $U_{n,m}$ as an example, $E_{n,m}$ denotes the incident electromagnetic wave, $\widetilde{E}_{n,m}$ represents the reflected electromagnetic wave, and $\mathit{\Gamma}_{n,m}$  is the reflection coefficient of the $U_{n,m}$ to perform the modulation and emission of the reflected wave. Due to the discontinuity in the air impedance of the transmission medium, $\widetilde{E}_{n,m}$ can be expressed as 

\begin{equation}
\widetilde{E}_{n,m} = \mathit{\Gamma_{n,m}} E_{n,m}, \quad n \in [1,N], m \in [1, M].
\end{equation}
$\mathit{\Gamma_{n,m}}$ is a complex parameter and can be denoted as

\begin{equation}{\mathit{\Gamma}}_{n,m} =  A _{n,m}e^{j\varphi _{n,m}}, \end{equation}
where $A_{n,m}$ and $\varphi_{n,m}$ represent the controllable amplitude
and phase shift of $U_{n,m}$.

In addition, the coefficient is linked to the equivalent load impedance $Z_{n,m}$ of the unit cell and the characteristic impedance of the air $Z_0$, which is written as \cite{r30}

\begin{equation}\mathit{\Gamma_{n,m}}=\frac{Z_{n,m}-Z_0}{Z_{n,m}+Z_0}.\end{equation}

By combining (5) and (6), the amplitude and the phase of the
reflection coefficient $\mathit{\Gamma_{n,m}}$ can be obtained as
\begin{equation}\color{blue}
    A_{n,m} = \left|\mathit\Gamma_{n,m}\right|,
\end{equation}

and
\begin{equation}\color{blue}
  \varphi_{n,m} = \arctan{\left(\frac{Im{(\mathit\Gamma_{n,m})}}{Re(\mathit\Gamma_{n,m})}\right)}.  
\end{equation}


 The load impedance $Z_{n,m}$  can be dynamically tuned through electronic control mechanisms. In the RIS-assisted signal transmission framework, the process begins with an initial signal generated by an excitation source. This signal carries a frequency of $f_c$ and an amplitude of $A_c$. Thus, the equation (4) can be further expressed as

\begin{equation}
\begin{aligned}
{\widetilde{E}}_{n,m}& = {A}_{n,m}{e}^{j{\varphi }_{n,m}}{A}_{c}{e}^{{j2\pi }{f}_{c}t} \\
&= {A}_{c}{A}_{n,m}{e}^{j\left( {{2\pi }{f}_{c}t + {\varphi }_{n,m}}\right) }. 
\end{aligned}
\end{equation}

\subsection{Path Gain}

\textcolor{blue}{In terahertz wireless communication, the spreading loss and molecular absorption loss are identified as the primary contributing factors\cite{r31}. According to the Friis formulation, the loss of free spreading is given by \cite{r32}}

\begin{equation} {L}_{\text{spread }}\left( {f_c,d}\right)  = {\left( \frac{4\pi f_cd}{c}\right) }^{2},\end{equation}
where $d$ is the spread distance, $c$ stands for the speed of light in free space. $L_\text{spread}$ is proportional to the square of the frequency, so the loss is more severe at terahertz frequencies. And the absorption loss that absorbed by water molecules in the atmosphere is denoted as\cite{r33}

\begin{equation} {L}_{\mathrm{{abs}}}\left( {f_c,{d}}\right) = {e}^{\alpha\left( f_c\right) {d}},\end{equation}
here, $\alpha(f_c)$ represents the molecular absorption loss coefficient, which is a frequency-dependent parameter. The absorption coefficient varies with different transmission media in the atmosphere, with the primary source of absorption loss being water vapor molecules. 

\subsection{Antenna Gain}

\begin{figure}
\centering
\includegraphics[width=1.6in,height=1.5in]{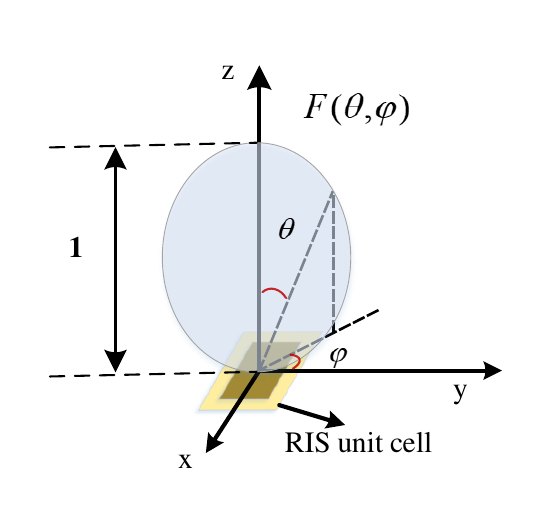}
\caption{Diagram of normalized power radiation pattern.}
\label{fig: radiated_pattern}
\end{figure}

The gain of the RIS unit cell describes the amount of power transmitted or received in the radiation direction relative to an isotropic antenna. Assuming 100$\%$ antenna efficiency, Let $G$ represent the gain of the RIS unit, which can be expressed as\cite{r34}

\begin{equation}G=\frac{4\pi}{\int_{\varphi=0}^{2\pi}\int_{\theta=0}^{\pi}F\left(\theta,\varphi\right)\sin\theta d\theta d\varphi},\end{equation}
here, $F(\theta,\varphi)$ represents the normalized radiation pattern of RIS unit cell in spherical coordinates, as shown in Fig. \ref{fig: radiated_pattern}, where $\theta$ and $\varphi$ denote the azimuth and elevation angles of the transmission direction, respectively. The power radiation pattern defines the variation of transmitted or received power in different directions and is expressed by equation (\ref{F})

\begin{equation}\left.F\left(\theta,\varphi\right)=\left\{
\begin{array}
{cc}{\mathrm{cos}\theta} & {\theta\in\left[0,\frac{\pi}{2}\right],\varphi\in[0,2\pi]} \\
{0} & {\theta\in\left[\frac{\pi}{2},\pi\right],\varphi\in[0,2\pi]}
\end{array}\right.\right..
\label{F}
\end{equation}

\subsection{Power Transmission}

The power of incident carrier signal into  the unit cell $U_{n,m}$ can be expressed as

\begin{equation}P_{n,m}^{in}=Sd_{x}d_{y},\end{equation}
where $S$ is the average power density, and the electric field of the incident carrier signal into $U_{n,m}$ is given by

\begin{equation}E_{n,m}^{in}=\sqrt{2Z_{0}S}e^{j2\pi f_{c}t}.\end{equation}

\begin{figure*}[bp]\setcounter{equation}{17}
\centering
\begin{align}
E_{n,m}^{r} &= \sqrt{\frac{2Z_0 P_{n,m}^{r}}{A_r}} e^{-j\left(\frac{2\pi d_{n,m}}{\lambda} + \varphi_{n,m}\right)} e^{j2\pi f_c t} \notag \\
&= \sqrt{\frac{2Z_0 G_{t} P_{n,m}^{in} F^{t}\left( {{\theta }_{n,m}^{t},{\varphi }_{n,m}^{t}}\right) {F}^{r}\left( {{\theta }_{n,m}^{r},{\varphi }_{n,m}^{r}}\right)}{4\pi d_{n,m}^2}} \mathit{\Gamma_{n,m}} e^{\frac{-j2\pi d_{n,m}}{\lambda}} e^{j2\pi f_c t} \notag \\
&= \frac{\sqrt{Z_0 G_{t} P_{n,m}^{in} F^{t}\left( {{\theta }_{n,m}^{t},{\varphi }_{n,m}^{t}}\right) {F}^{r}\left( {{\theta }_{n,m}^{r},{\varphi }_{n,m}^{r}}\right)}}{\sqrt{2\pi} d_{n,m}} e^{\frac{-j2\pi d_{n,m}}{\lambda}}  A_{n,m} e^{j\varphi_{n,m}} e^{j2\pi f_c t}. 
\end{align}
\end{figure*}

\begin{figure*}[bp]\setcounter{equation}{21}
\centering
\begin{align}
P_r &= \frac{G_t G_r \lambda^2 }{16\pi^2} \left( \sum_{m=1}^{M} \sum_{n=1}^{N} \frac{\sqrt{P_{n,m}^{in} F^{t}\left( {{\theta }_{n,m}^{t},{\varphi }_{n,m}^{t}}\right) {F}^{r}\left( {{\theta }_{n,m}^{r},{\varphi }_{n,m}^{r}}\right)}}{d_{n,m}} e^{\frac{-j2\pi d_{n,m}}{\lambda}}\mathit{ \Gamma_{n,m} }e^{j2\pi f_c t} \right)^2.
\end{align}
\end{figure*}

According to the law of energy conservation, the power of reflected signal by the unit cell $U_{n,m}$ is determined as the product of the incident signal power and the square of the reflection coefficient. This relationship can be written as
\setcounter{equation}{15}
\begin{equation}{P}_{n,m}^{in}\left| {\Gamma }_{n,m}^{2}\right|  = {P_{n,m}^{t}},\end{equation}
$P_{n,m}^t$ represents the power of reflected signal by the transmitter.

Considering ${L}_{\text{spread}}$ and ${L}_{\text{abs}}$, the power of the reflected signal received by the receiver from the reflection of the unit cell $U_{n,m}$ of the transmitter can be expressed as\cite{r35}

\begin{equation} \setcounter{equation}{17}
{P}_{n,m}^{r} = \frac{P_{n,m}^{t}G_{t}G_{r}F^{t}\left( {{\theta }_{n,m}^{t},{\varphi }_{n,m}^{t}}\right) {F}^{r}\left( {{\theta }_{n,m}^{r},{\varphi }_{n,m}^{r}}\right)  }{{L}_{\text{spread }}\left( {f_c,d_{n,m}}\right) {L}_{\text{abs }}\left( {f_c,d_{n,m}}\right)}, 
\end{equation}
where $G_t$ and $G_r$ denote the gains of the transmitter and receiver, respectively, while ${F}^{t}\left( {{\theta }_{n,m}^{r},{\varphi }_{n,m}^{r}}\right)$is the power normalization function for the transmitter, and ${F}^{r}\left( {{\theta }_{n,m}^{r},{\varphi }_{n,m}^{r}}\right)$ is the power normalization function for the receiver, $d_{n,m}$ is the distance between unit cell $U_{n,m}$ and receiver. 

Combining equations (15), (16), and (17), the electric field by the receiver from unit cell $U_{n,m}$ can be obtain as equation (18), where $e^{-j\left(\frac{2\pi d_{n,m}}{\lambda} + \varphi_{n,m}\right)}$ is phase shift introduced by propagation and reflection coefficient of the $U_{n,m}$. The total electric field of the receiver is the summation of the electric fields reflected by all the unit cell towards it, which can be expressed as

\begin{equation}\setcounter{equation}{19}
{ E}_{r} = \mathop{\sum }\limits_{{m = 1}}^{M}\mathop{\sum }\limits_{{n = 1}}^{N}{E}_{n,m}^{r}. 
\end{equation}

Furthermore, the relationship between the received signal power and the total electric field can be represented as

\begin{equation} \setcounter{equation}{20}
{P}_{r} = \frac{{E}_{r}^{2}}{2{Z}_{0}}{A}_{r}, \end{equation}
and $A_r$ is the aperture of receiver can be written as 
\begin{equation}\setcounter{equation}{21}
A_r=\frac{G_r\lambda^2}{4\pi}.
\end{equation}

Finally, by combining (19), (20) and (21), the power of the receiver array can be obtained as shown in equation (22).

\subsection{Power Cycle Establishment}

\textcolor{blue}{In the power cycle establishment, the received power at the UE is split by the PD, with the majority used for UE operation and the remainder fed back to the BS for amplification to sustain the power cycle. During the initial iterations, the transmitted signal undergoes significant spatial spreading and path loss. The amplification gain is insufficient to compensate for the propagation loss, resulting in low received power at the UE. As the number of iterations increases, the retro-reflected signals traveling along equidistant paths maintain consistent phase offsets. This enables progressive phase alignment through coherent superposition at both the BS and the UE, leading to a narrowing of the beamwidth and thereby enhancing energy concentration in the target direction. Eventually, the system reaches a stable state in both power and phase.}

\begin{figure}\color{blue}
\centering                 
\includegraphics[width=2.8in]{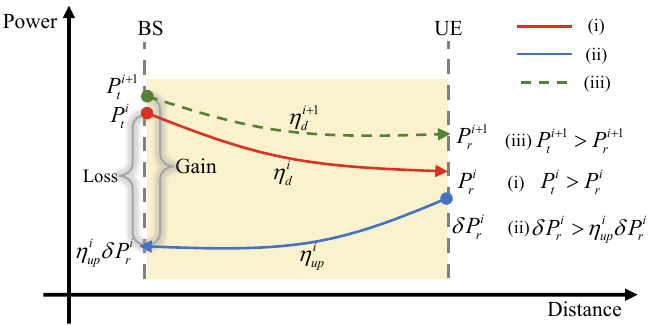}
\caption{Diagram of power variation in iteration.}
\label{fig: Power cycle}
\end{figure}

\textcolor{blue}{The power variation during the iteration process is illustrated in Fig. \ref{fig: Power cycle}. At the $i-th$ iteration, the BS transmits with power $P_{t}^{i}$ over the downlink with efficiency $\eta_{d}^i$, resulting in a received power $P_{r}^i$ at the UE, as shown by (i) in the Fig. 6. Then, the UE captures the majority of received power for charging and decoding, while reflecting a portion of the power (denoted by ration $\delta$) back to the BS via the uplink with efficiency $\eta_{up}^i$, as illustrated in Fig. 6(ii). Finally, the BS amplifies the received power using gain function, which serves as the transmit power  $P_{t}^{i+1}$ for the $i+1-th$ iteration. This process corresponds to Fig. 6(iii). The underlying theoretical relationship is described by equation (23).}
    
\begin{equation}\setcounter{equation}{23}
\begin{cases}
P_{r}^{i} = \eta_{d}^{i}P_{t}^{i}\\
P_{t}^{i+1}=g(\eta_{up}^{i}\delta P_{r}^{i})
\end{cases}
\label{power cycle}
\end{equation}
\textcolor{blue}{where $P_{t}^{i}$ and $P_{r}^{i}$ represent the BS transmit power and UE received power in the $i-th$  iteration, respectively, and $g$ denotes the power amplification function at the BS. The BS transmit power $P_{t}^{i+1}$ in the $i+1-th$ iteration is derived from the retro-reflected power from the UE,  $P_{r}^i\cdot \delta$, multiplied by the uplink transfer efficiency $\eta_{up}^i$, and then amplified by $g$. To facilitate understanding, a specific numerical example is presented below. Let the initial power for the $i-th$ iteration be 1 mW, with $\eta_{d}^i$ = $\eta_{up}^i$ = 90\%, and $\delta$ = 1\%, the power amplification function is $2 \cdot P_{in}$. Thus, the power received at the UE from the BS is $P_{r}^i = 0.9 \text{ mW}$, according to the splitting ratio $\delta$, the portion of power returned to the BS is 0.009 $\text{ mW}$. After amplification with the efficiency $\eta_{\text{up}}^i$, the power received by the BS in the next iteration $P_{r}^{i+1}$ is 0.0162 $\text{ mW}$. }

\textcolor{blue}{Therefore, in the $i-th$ iteration, the path loss and power gain can be expressed by equation (24).}
    
\begin{equation}
\begin{cases}
\mathcal{L}oss_i=P_t^{i}-\eta_{up}^{i}\delta P_r^{i}=(1-\delta\eta_{up}^{i}\eta_{d}^{i})P_t^{i}\\
\mathcal{G}ain_i=P_{t}^{i+1}-\eta_{up}^{i}\delta P_r^{i}
\end{cases}
\end{equation}
\textcolor{blue}{where, the path loss represents the difference between the transmit power and the received power at BS during that iteration, while the power gain is defined as the difference between the amplified power and the received power at the BS.}

For the power cycle mechanism to sustain before reaching steady-state, the $\mathcal{G}ain_{i}$ at the $i-th$ iteration should always be greater than the $\mathcal{L}oss_{i}$, as the number of iterations increases $i\to i_\infty$, the $\mathcal{L}oss_{i}$ and $\mathcal{G}ain_{i}$ asymptotically converge to equality, indicating that the gain precisely compensates for the link loss. Thus, the received power stabilizes and no longer changes, confirming the successful establishment of the power cycle model. The convergence criterion in the power cycle model is mathematically formalized in equation (\ref{power_cycle_equation}).

\begin{equation}\begin{cases}
\mathcal{G}ain_1>\mathcal{L}oss_1 \\
\mathcal{G}ain_{i\to\infty}=\mathcal{L}oss_{i\to\infty} & 
\end{cases}
\label{power_cycle_equation}
\end{equation}
where $\mathcal{G}ain_1$ and $\mathcal{L}oss_1$ are the gain and loss in the $1-th$ iteration.

\bluesection{SWIPT MODEL}

\textcolor{blue}{This section analyzes the SWIPT performance of the proposed system, outlining the roles of the PD and rectifier in power diversion and conversion, and characterizes the received signal, signal-to-noise ratio (SNR), and capacity.}

\subsection{Charging Power for THz Rectifier}

\textcolor{blue}{The UE architecture employs the PD to split the SWIPT power between energy transmission and information transmission via a proportional splitting mechanism in each time slot.  It can be expressed as}

\begin{equation}
\begin{cases}
    P_\text{e} +P_\text{i} = (1-\delta) P_{{r}}\\
    P_\text{i} = \gamma(1-\delta ) P_{{r}}\\
    P_\text{e} = (1-\gamma)(1-\delta) P_{{r}}
\end{cases}
\end{equation}
\textcolor{blue}{where $1-\delta$ is the ratio allocated to SWIPT, $P_{{e}}$ and $P_{{i}}$ are the received power for energy and information, respectively, $\gamma\in[0,1]$ represents the information-centric power-splitting ratio.} The receiver antenna incorporates integrated power rectification circuit, converting the $P_\text{e}$ into stabilized DC output. For precise characterization of the rectification process, a designed GaN Schottky barrier diode model is implemented \cite{r36}, enabling calculation of the end-to-end power conversion efficiency $\eta_\text{rect}$ across the sub-terahertz band, which is
\begin{equation}
    P_\text{ch} = \eta_\text{rect}P_\text{e},
\end{equation}
where $P_\text{ch}$ denotes the charging power through the rectifier, and $\eta_\text{rect}$ 
represents the diode conversion efficiency.

\subsection{Communication Channel}

 \textcolor{blue}{The proposed system modulates signals by leveraging RIS unit cells to adjust phase and amplitude, bypassing the need for traditional per-unit RF chains. The carrier wave from the feed antenna is modulated in air using schemes such as QPSK \cite{r37} or 8PSK \cite{r38} and then reflected to the receiver. Communication is strategically initiated after the power cycle mechanism reaches stability, thereby minimizing multipath interference and confining transmission to a focused path for superior signal quality. The corresponding baseband received signal in noise follows equation (\ref{channel}).}

\begin{equation}y(t)=\sum_{m=1}^{M}\sum_{n=1}^{N}\sqrt{P_{n,m}^{in}}h_{n,m}A_{n,m}e^{j\varphi_{n,m}}e^{j2\pi f_{c}t}+n(t),
\label{channel}
\end{equation}
and

\begin{equation}\begin{aligned}
h_{n,m}\\
 & =\frac{\sqrt{G_tG_r\lambda^2F^t\left(\theta_{n,m}^{t},\phi_{n,m}^{r}\right)F^{r}\left(\theta_{n,m}^{r},\phi_{n,m}^{r}\right)}}{4\pi d_{n,m}} \\
 & \times e^{\frac{-j2\pi d_{n,m}}{\lambda}}{},
\end{aligned}\end{equation}
where $y(t)\in\mathbb{C} ^{1\times NM}$ and $h=[h_{1,1},h_{1,2},...,h_{n,m}]\in\mathbb{C} ^{1\times NM}$ are the received signal vector and the wireless channel matrix between BS and UE, $n(t)$ represents the additional Gaussian white noise added at the UE, modeled as $\mathcal{CN}\left(\mathbf{0}_\text{N},\sigma_\text{r}^{2}\mathbf{I}_\text{N}\right)$.

\textcolor{blue}{In addition, the power-consuming amplification process generates thermal noise due to active component imperfections, modeled as  $\mathcal{CN}\left(\mathbf{0}_\text{N},\sigma_\text{a}^{2}\mathbf{I}_\text{N}\right)$\cite{r39}. Since both signal and noise are amplified, the total noise power equals the sum of their variances,}

\begin{equation}
    \begin{aligned}
        \sigma_\text{tot}^{2}   = G(\sigma_\text{r}^{2}+\sigma_\text{a}^{2}+\sigma_\text{p}^{2}),
    \end{aligned}
\end{equation}
where $\sigma_\text{tot}^{2}$ represents the total output noise variance, $\sigma_\text{p}^{2}$ characterizes the additive noise variance introduced by the amplifier. Then the SNR can be depicted as

\begin{equation}
    \text{SNR} = \frac{\gamma P_t}{\sigma ^{2}_\text{tot} }.
\end{equation}

The channel capacity of our system then be obtained as
 \begin{equation}
 \text{C}=\text{B}\log_2 ( 1+\mathrm{SNR} ),\end{equation}
where B is the system bandwidth.

\section{NUMERICAL RESULTS AND ANALYSIS}

\textcolor{blue}{To evaluate the performance of the proposed RIS-assisted THz self-alignment SWIPT system, this section examines the evolution of power distribution during the establishment of the self-alignment wireless link, analyzes the trends in both charging power and transfer efficiency, and investigates the influence of the FoV on system performance. For comparison, we also evaluate a unidirectional conjugate system that relies solely on UE conjugation, the proposed bidirectional conjugate architecture achieves higher transfer efficiency and received power. The transfer efficiency at the $i-th$ iteration is defined as the ratio of the power received at the UE to the  power transmitted by the BS and is given by}

\begin{equation}\color{blue}
   \eta^{i}_{d} = \frac{P^{i}_{r}}{P^{i}_{t}},
\end{equation}
\textcolor{blue}{where $i$ represents the iteration times, as $i$ approaches infinity, that is the system reaches a steady state, the resulting value represents the final transfer efficiency.}

\subsection{Parameters Setting}

\begin{table}[htbp]\color{blue}
    
    \centering
    \caption{system parameters}
    \renewcommand{\arraystretch}{1.2}
    \begin{tabular}{ccc}  
    \toprule  
    \textbf{Parameter} & \textbf{Symbol} & \textbf{Value} \\ 
    \midrule  
    Carrier frequency & $f_c$ & 135 GHz \\
    Wavelength & $\lambda$ & 2.2 mm \\
    System bandwidth & $B$ & 20 GHz \\
    Spacing of unit cells & ${d_x},{d_y}$ & $\frac{\lambda}{2}$ \\
    Amplitude response of RIS & $A_{n,m}$ & 0.95 \\
    Amplifier gain function & $G(\cdot)$ & 21.9 dB (Max)\cite{r40} \\
    Gain for antenna and RIS & $G_r$,$G_t$ & $\pi$ (Max) \cite{r41}\\
    Molecular absorption loss & $k$ & $9.217\times 10^{-4}$ dB/m \cite{r42}\\
    PA noise figure & $F_p$ & 8.5\, \text{dB} \cite{r45} \\
    Characteristic impedance & $Z_0$ & 120 $\pi$ \\
    Ambient temperature & ${T}$ & 300 K \\
    Boltzmann constant & $k_B$ & $1.38\times10^{-23}$ \\
    PA noise  & $\sigma _{p}^2$ & $2Z_{0}k_{B}kTF_{p}B$  \\
    Antenna noise &   $\sigma _{r}^2$ & -70 dBm  \cite{r44}\\
    Active RIS noise  & $\sigma _{a}^2$ & -70 dBm  \cite{r44} \\
    Conversion efficiency & $\eta_{\text{rect}}$ & 78\%\cite{r36} \\
    Power ratio for demodulation & $\gamma$ & $5\%$ \\
    Power ratio for feedback & $\delta$ & $1.5\%$ \\

    \bottomrule  
    \end{tabular}
    \label{tab1}
\end{table}

\textcolor{blue}{The system parameters are summarized in Table \ref{tab1}. The amplifier architecture follows the hybrid linear-nonlinear design framework established in \cite{r40}, achieving a peak gain of 21.9 dB through coordinated operation of its cascaded linear and nonlinear stages. The RIS unit cell gain radiation pattern, adopted from \cite{r41}, assuming that the BS and UE share the same gain. Molecular absorption loss coefficient is 9.217 × 10$^
{-4}$ dB/m at the operational frequency\cite{r42}. The RIS and receiver antenna are uniform planar arrays with an half-wavelength spaced unit cells \cite{r43}. And the noise $\sigma_\text{r}^2$ = $\sigma_\text{a}^2$ = $-$70 dBm \cite{r44}, $\sigma_{p}^2$ = 2$Z_{0}k_BT_{o}F_{p}B$, $k_B$ is the Boltzmann constant. For geometric system modeling, the RIS plane is embedded in a Cartesian coordinate system with its phase center coinciding with the origin (0, 0, 0) of the xOy-plane. The UE is positioned along the positive z-axis under boresight alignment, maintaining a line of sight propagation distance $d$ orthogonal to the RIS plane. Assuming an initial system power of 1 mW, the signal is initially scattered by the RIS.}

\begin{figure*}\color{blue}
\centering
\includegraphics[width=6.6in,height=1.8in]{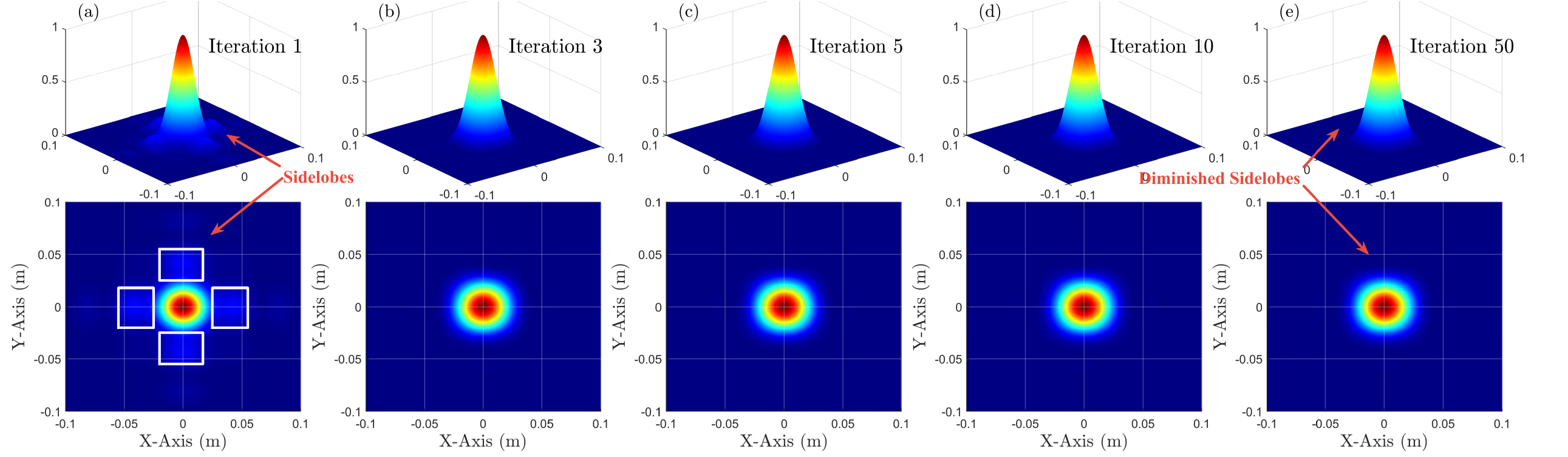}
\caption{Evolution of normalized power distribution at the receiver throughout the iterative optimization process. Subplots (a)-(e) correspond to iteration counts of 1, 3, 5, 10, and 50, respectively.}
\label{fig: space-distribution}
\end{figure*}

\subsection{Self-alignment Performance Analysis}

\textcolor{blue}{To illustrate the self-alignment process during the iteration propagation of the terahertz signal, Fig. \ref{fig: space-distribution} shows the dynamic evolution of the normalized power density distribution across a 60 × 60 RIS array at the UE side, located at a distance $d$ = 1 m, during the 1st, 3rd, 5th, and 50th iterations. The normalized power distribution (NPD) is defined by equation (\ref{NPD}), where $P_r(x,y)$ denotes the computed power density at coordinates (x,y) on the UE, and $max(P_r(x,y))$ represents the maximum value of the power distribution. In the two-dimensional distribution plots, the plane pattern represents the power distribution along the center column of the UE, and the values on both sides of the x-axis represent the sidelobe effect of the UE array surface. The Fig. \ref{fig: space-distribution}(a) shows the initial state at this point, the signal from BS has not yet been fully aligned with the UE. Therefore, the energy distribution is relatively scattered, with obvious side lobes protruding. As the number of iterations increases, as shown in Fig. \ref{fig: space-distribution}(b)-(e) the energy distribution gradually becomes concentrated, with sidelobes substantially suppressed and the signal accurately directed toward the UE. This improvement results from the gradual phase alignment and coherent superposition of signals during conjugate reflection, ultimately reaching a steady state. The system adaptively calibrates the unit phases at both the BS and UE, effectively suppressing sidelobe artifacts and establishing a self-alignment wireless transmission link, effectively overcoming beam misalignment.}

\begin{equation}\color{blue}
   \mathrm{NPD} = =\frac{P_r(x,y)}{\max{(P_r(x,y))}}.
    \label{NPD}
\end{equation}

\begin{figure}[ht]\color{blue}
    \centering
    \includegraphics[width=0.8\linewidth, height=0.2\textheight]{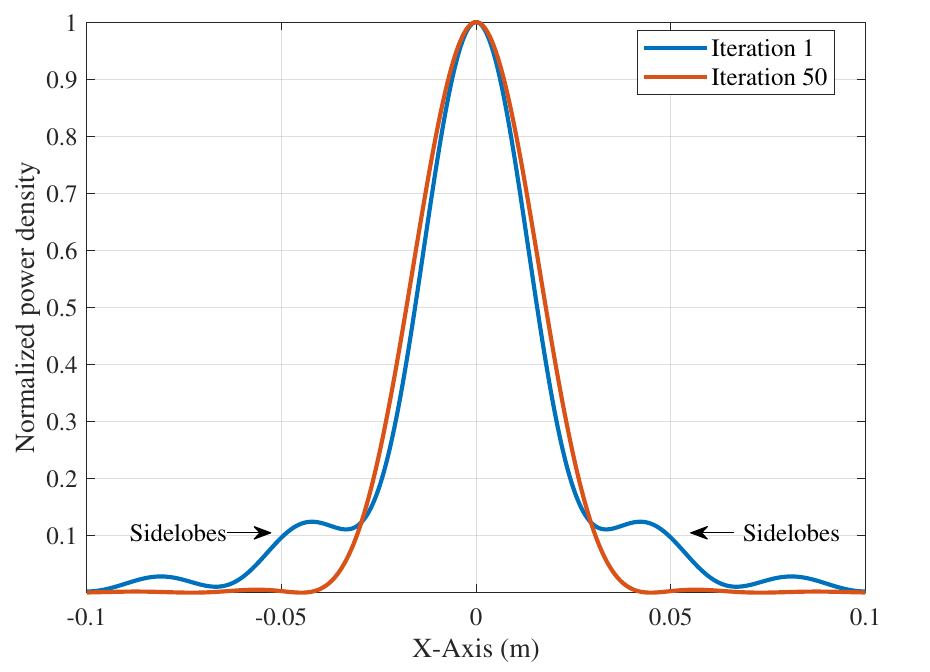}
    \caption{Comparative analysis of receiver power density evolution: from nascent stage (iteration 1) to convergence state (iteration 50).}
    \label{fig:normized_side}
\end{figure}

\textcolor{blue}{Fig. \ref{fig:normized_side} provides a direct comparison of the sidelobe effects, where the two curves represent the two-dimensional data profiles along the central row of the UE corresponding to the 1st and the 50th iterations shown in Fig. \ref{fig: space-distribution}, more intuitively demonstrating the gradual suppression of sidelobes as the number of iterations increases. An iteration of the system corresponds to the complete transmission process of the signal from the BS to the UE and back, reflected back to the BS. This process also represents one full computation of the power cycle model. During the initial iteration, significant power leakage leads to noticeable sidelobe effects at the center of the receiver. Upon reaching the steady state through iterations, the main lobe converges to a fixed point with negligible sidelobes, indicating a substantial reduction in sidelobe effects. The signal becomes concentrated in the direction of maximum power at the UE, with the beam accurately focused toward the UE.}

\begin{figure}[ht]\color{blue}
    \centering
    \includegraphics[width=0.8\linewidth, height=0.2\textheight]{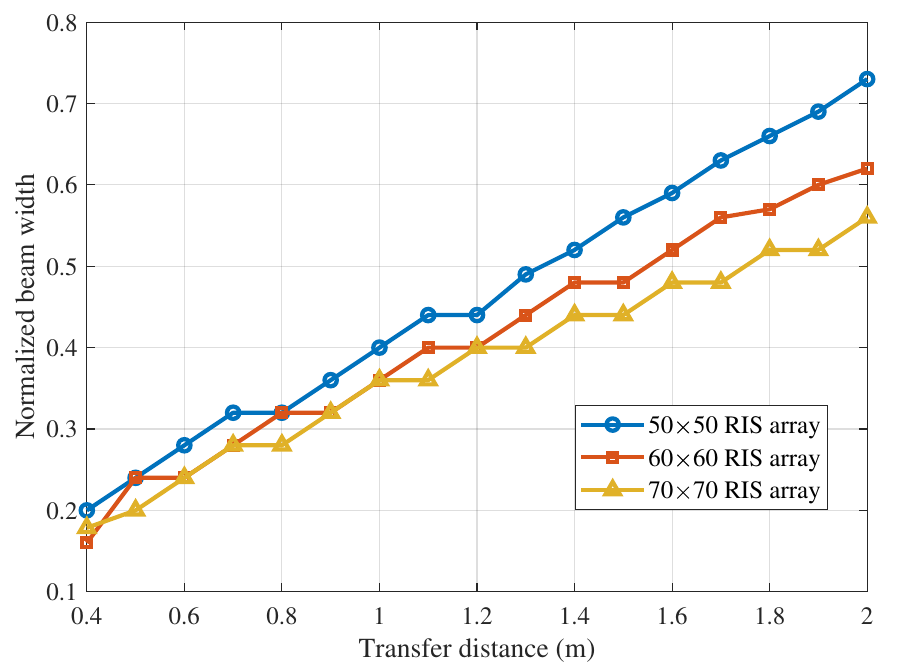}
    \caption{Characterization of beam widths across RIS array architectures at half power.}
    \label{fig:beam_width}
\end{figure}

\begin{figure*}[t]\color{blue}
    \centering
    \includegraphics[width=6.6in,height=2.5in]{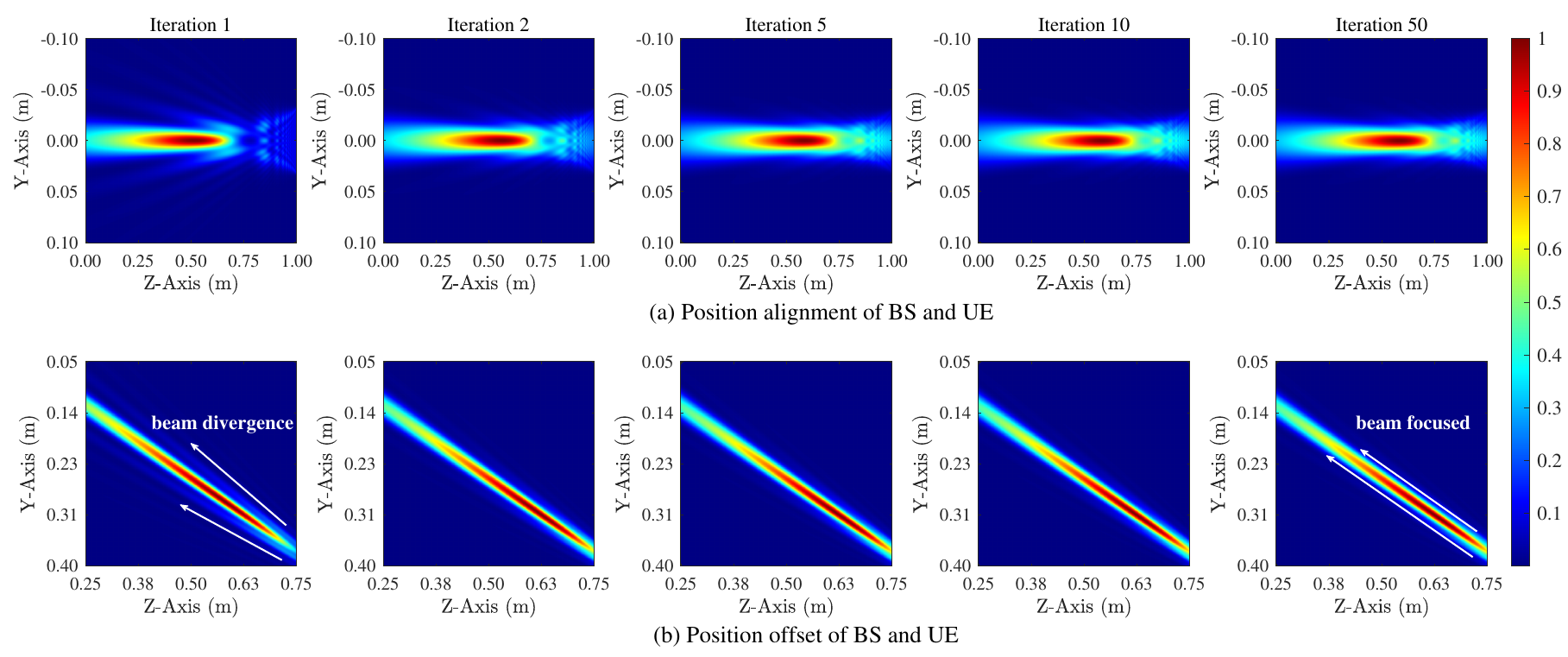}
     \caption{Evolution of the spatial power distribution and beam focusing at iterations 1, 3, 5, 10, and 50 for a 60 × 60 array system with a link distance of $d$ = 1 m and a 0.5 m offset of the UE side along the x-axis.}
    \label{fig:BS_UE_distribution_power_density}
\end{figure*}

\textcolor{blue}{To clarify the beam focusing capability of the proposed system, Fig. \ref{fig:beam_width} shows the beam spot width at the UE under steady state for different array dimensions and transfer distances. The normalized beam width (NBW) is defined as the angular range over which the radiated power decreases to half of its maximum value in the main lobe, as shown in equation (\ref{NBW}), where $\Delta x=\arg\min_x\left|P_r(x,y_{mid})-\frac{max(P_{r}(x,y_{mid}))}{2}\right|-x_{\mathrm{max}}$, here, $P_r(x,y_{mid})$ denotes the power at position index $y$ along the center row of the UE array, $\frac{max(P_{r}(x,y_{mid}))}{2}$ is the half-power point, $x_{\mathrm{max}}$ is the position index corresponding to the maximum power along the center row, and $L$ is a normalization length factor defined as the distance from the maximum power position to the end of the array. A smaller spot width indicates a more concentrated energy. Numerical results demonstrate an approximately linear beam broadening progression as transfer distance increases. This phenomenon arises from progressive wavefront divergence during extended propagation, which degrades spatial energy confinement. And a inverse proportionality exists between array scale and beam width at fixed operational distances. Comparative analysis reveals that larger arrays (N = 70) achieve 21.4$\%$ tighter beam confinement than their smaller counterparts (N = 50) at $d$ = 1.5 m. This dimensional dependence stems from enhanced energy focusing capability enabled by expanded effective aperture areas in scaled arrays, which effectively mitigates wavefront distortion induced by path loss.}

\begin{equation}\color{blue}
    \mathrm{NBW} = 2\cdot\frac{\Delta  x }{L}.
    \label{NBW}
\end{equation}

 \textcolor{blue}{Fig. \ref{fig:BS_UE_distribution_power_density} illustrates the spatial radiation patterns of the signal transmitted from the UE to the BS for the proposed system with a 60 × 60 array at a transfer distance of $d$ = 1 m and an offset of 0.5 m along the x-axis at the UE, corresponding to the 1st, 2rd, 5th, and 50th iterations. To better visualize the beam variation, a magnified partial view is provided, where the horizontal axis represents the transfer distance ranging from 0.25 m to 0.75 m, and the vertical axis indicates the distance from the center of the UE planar array. As shown in Fig. \ref{fig:BS_UE_distribution_power_density} (a), it can be observed that during the initial iteration (iteration 1), the spatial distribution of the signal is relatively diffuse, failing to concentrate effectively within the communication link. As the number of iterations increases, the signal undergoes coherent superposition through conjugate reflection, leading to gradual phase alignment. Thus, the energy is concentrated along the transmission path, ultimately forming a highly focused communication link. This self-alignment property effectively mitigates the impact of beam misalignment. Moreover, when the UE position changes, the beam alignment is automatically re-established, thereby overcoming pointing deviations caused by relative movement between the BS and the UE, as depicted in Fig. \ref{fig:BS_UE_distribution_power_density}(b). }


\subsection{SWIPT Performance Analysis}

\begin{figure}[h]\color{blue}
\centering
\includegraphics[width=0.8\linewidth, height=0.2\textheight]{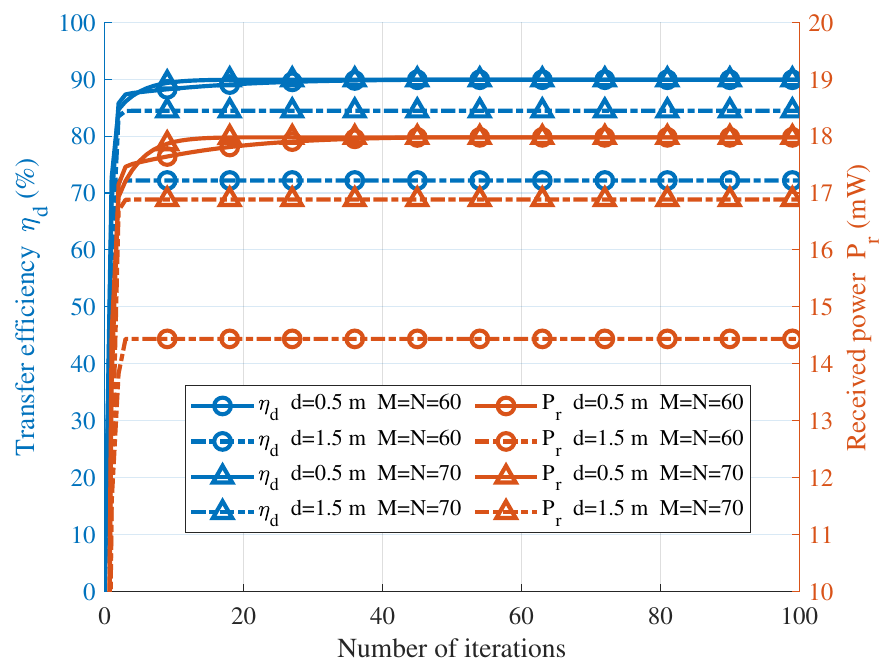}
\caption{Convergence dynamics of RIS arrays transfer efficiency \(\eta_d\) and received power \(P_r\) at \(d=0.5\,\mathrm{m}\) and \(d=1.5\,\mathrm{m}\).}
\label{fig:iteration}
\end{figure}


\textcolor{blue}{Fig. \ref{fig:iteration} illustrates the convergence behavior of transmission efficiency and received power during the self-alignment process. Results are shown for 60 × 60 and 70 × 70 RIS arrays at transfer distances of $d$ = 0.5 m (solid lines) and $d$ = 1 m (dashed lines). Initially, both power and efficiency remain near zero due to unoptimized array phases and significant path loss. As the number of iterations increases, both metrics exhibit a monotonic improvement until eventually converging to stable values, indicating the establishment of a steady power cycle. As expected, larger array dimensions yield better performance: the 70 × 70 RIS array achieves a peak efficiency of 84.43\% and a received power of 16.88 mW at $d$ = 1.5 m, representing an improvement of 16.96\% in efficiency and 16.98\% in power compared to the 60 × 60 array.}

\begin{figure}[h]\color{blue}
\centering
\includegraphics[width=0.8\linewidth, height=0.2\textheight]{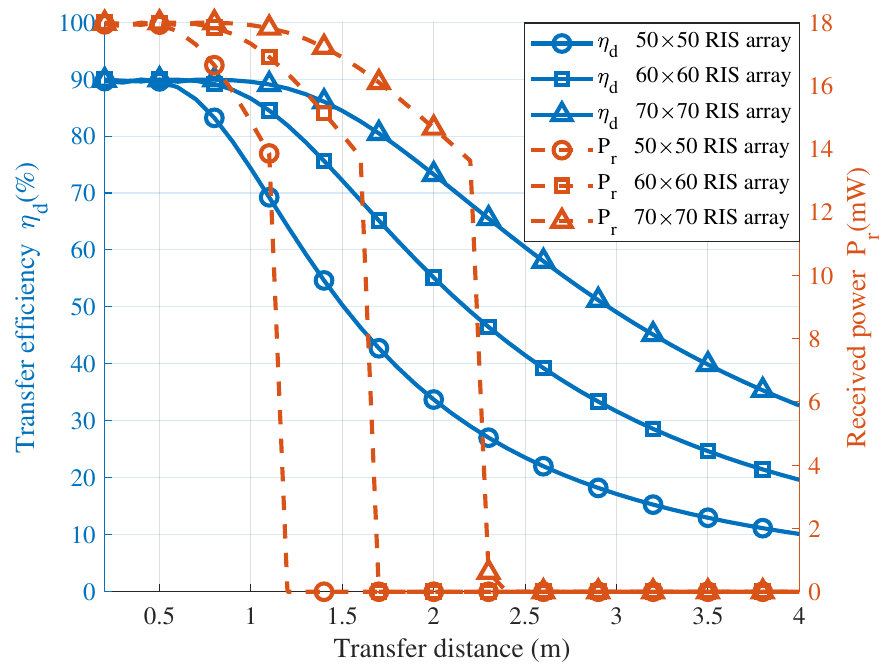}
\caption{Changes of transfer efficiency \(\eta_d\)and received power \(P_r\) with transfer distance under different size of RIS array.}
\label{fig:eff_po_distance}
\end{figure}

\textcolor{blue}{Fig. \ref{fig:eff_po_distance} presents an analysis of the maximum transfer distance under steady-state system operation, defined as the distance at which the radiated power approaches zero, for various RIS array sizes. Both the transfer efficiency and the received power are observed to decrease gradually as the transfer distance increases. Notably, the 70 × 70 RIS array consistently provides the highest received power, followed by the 60 × 60 and 50 × 50 arrays, indicating that larger RIS arrays provide superior beamforming capabilities and thus maintain better signal quality over extended distances. A notable feature is the abrupt decline in received power beyond a specific distance. As the transfer distance increases, the path loss escalates accordingly. At the UE, 1.5\% of the received power is used to sustain the resonance mechanism. This fraction is fed back to the BS and amplified; however, the amplified power remains relatively low. Consequently, the gain achieved through amplification becomes insufficient to offset the path loss, resulting in a sharp drop in the received power at the destination. Specifically, for RIS configurations with 50 × 50, 60 × 60, and 70 × 70 elements, the maximum transfer distances $d_{max}$ are found to be 1.1 m, 1.6 m, and 2.2 m, respectively. For the 70 × 70 array, the received power at the maximum distance of 2.2 m reaches 13.62 mW, corresponding to a transmission efficiency of 68.1\%.}

\begin{figure}[h]\color{blue}
\centering
\includegraphics[width=0.8\linewidth, height=0.2\textheight]{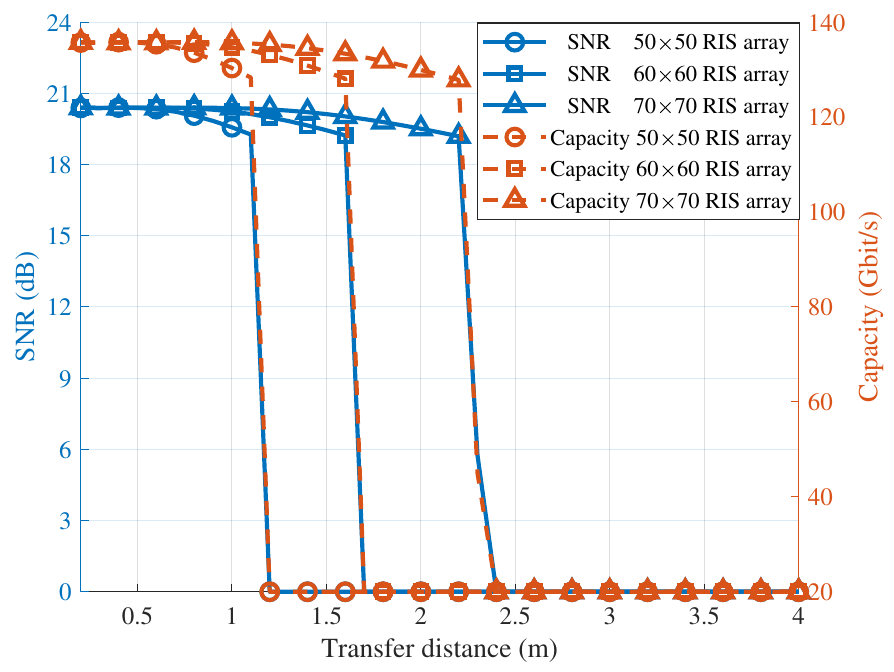}
\caption{Changes of SNR and Capacity with transfer distance under different size of RIS array.}
\label{fig:SNR_C_distance}
\end{figure}

\textcolor{blue}{Subsequently, the communication performance of the system is evaluated in Fig. \ref{fig:SNR_C_distance}. For all three array configurations, both metrics exhibit only a gradual decline within a certain range, with the degradation trend remaining relatively mild. Beyond critical distances of 1.1 m, 1.6 m, and 2.2 m for the respective arrays, the SNR and capacity stabilize at a low level. This behavior is consistent with the analysis of Fig. \ref{fig:eff_po_distance}, once the transfer distance exceeds these thresholds, the amplification gain can no longer compensate for the path loss, resulting in a sharp drop in received power. Consequently, the SNR and channel capacity also degrade significantly due to their strong dependence on received power. As expected, larger arrays demonstrate superior performance, achieving higher SNR and capacity at the same distance. For instance, at $d$ = 1.5 m, the 70 × 70 RIS array achieves an SNR of 20.12 dB and a capacity of 134 Gbit/s, outperforming the 60 × 60 elements configuration (19.44 dB, 129 Gbit/s). At the maximum transfer distance $d_{\text{max}}$ = 2.2 m, the 70 × 70 RIS array maintains a capacity of 127.84 Gbit/s with an SNR of 19.19 dB.}



\begin{figure}[h]\color{blue}
\centering
\includegraphics[width=0.8\linewidth, height=0.2\textheight]{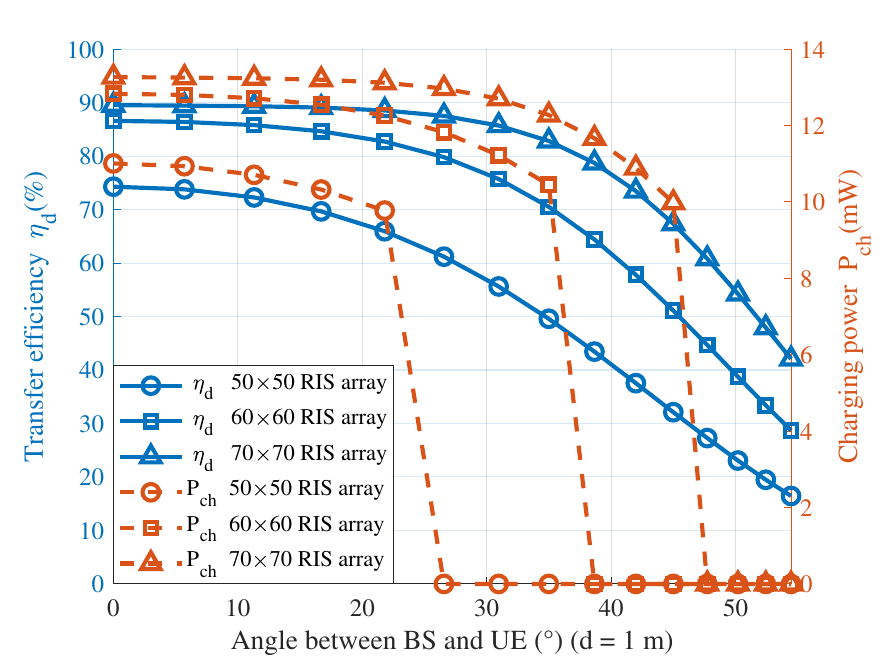}
\caption{Performance of transfer efficiency $\eta_d$ and charging power $P_{ch}$ as a function of angle between the BS and the UE.  }
\label{fig:eff_po_shift_1m}
\end{figure}

\textcolor{blue}{The influence of the FoV on both transfer efficiency and charging power is analyzed, as shown in Fig. \ref{fig:eff_po_shift_1m}. The horizontal axis represents the relative angular offset between the BS and the UE at a fixed collimated transfer distance of $d$ = 1 m. It can be observed that both the charging power and transfer efficiency gradually decrease as the angular misalignment between the BS and UE increases. A further increase in angular deviation leads to a sharp decline in charging power. This occurs because, at larger offset angles, the radiation gains of both BS and UE elements decrease significantly. Even after signal amplification, the path loss cannot be compensated, resulting in the received power approaching zero. Furthermore, although the charging power drops to nearly zero, the transfer efficiency remains at a non-negligible level due to the continuous operation of the steady-state power cycle mechanism. Notably, larger array sizes yield higher transfer efficiency and charging power, corresponding to a wider FoV. For instance, with a 70 × 70 array, the charging power delivered to the UE exceeds 9.98 mW, and a transfer efficiency of over 67.35$\%$ is maintained within a 90{°} FoV ($\pm $45{°}). And the maximum FoV with array dimension of 50 × 50 is 44{°}, within the FoV, $\eta_d$ and $P_c$ exhibit minimal variation ($\bigtriangleup \eta <$ 8.34$\%$, $\bigtriangleup P_c<$ 1.23 mW). }

\begin{figure}[h]\color{blue}
\centering
\includegraphics[width=0.8\linewidth, height=0.2\textheight]{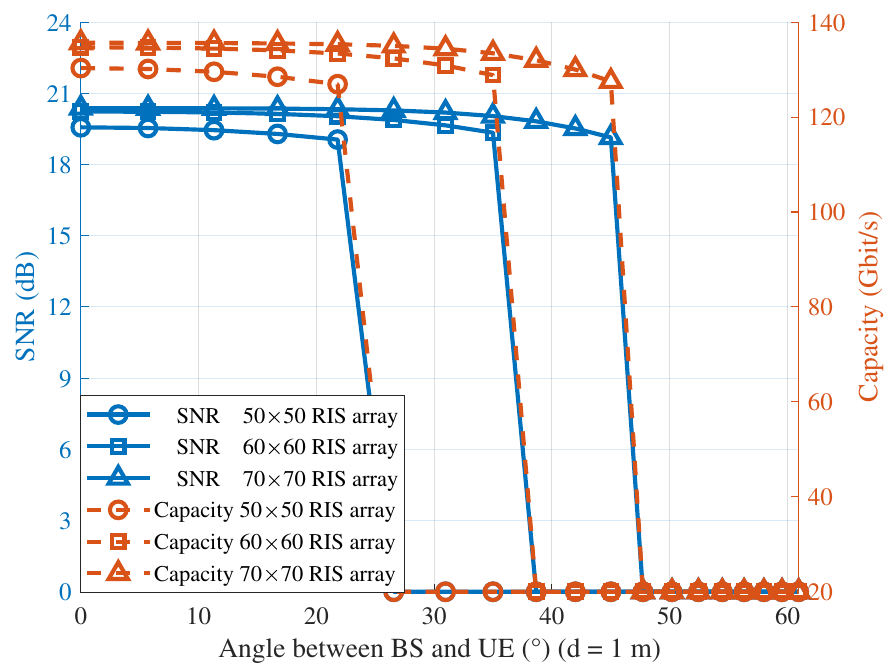}
\caption{ Performance of SNR and Capacity as a function of angle between the BS and the UE. }
\label{fig:SNR_C_shift_1m}
\end{figure}

\textcolor{blue}{Similarly, the variations in SNR and communication capacity under angular offset at $d$ = 1 m are analyzed and presented in  Fig. \ref{fig:SNR_C_shift_1m}. Within the FoV, the SNR and capacity remain at high values with only a slight degradation. However, as the angular offset further increases, the radiation gain of the system decreases, causing the received power to drop to a negligible level. This results in a sharp decline in both SNR and channel capacity, significantly deteriorating the signal quality of the communication link. Moreover, larger array sizes exhibit greater tolerance to angular misalignment. In addition, both SNR and channel capacity improve as the number of elements in the RIS array increases. Compared to the 60 × 60 and 50 × 50 configurations, the 70 × 70 RIS array consistently delivers higher SNR and capacity across the entire angular range. At the maximum allowable angular offset for 70 × 70 array, the system achieves an SNR of approximately 19.14 dB and a channel capacity of about 127 Gbit/s.}

\subsection{Experimental Comparison Performance Analysis}

\begin{figure}[h]\color{blue}
\centering
\includegraphics[width=0.8\linewidth, height=0.2\textheight]{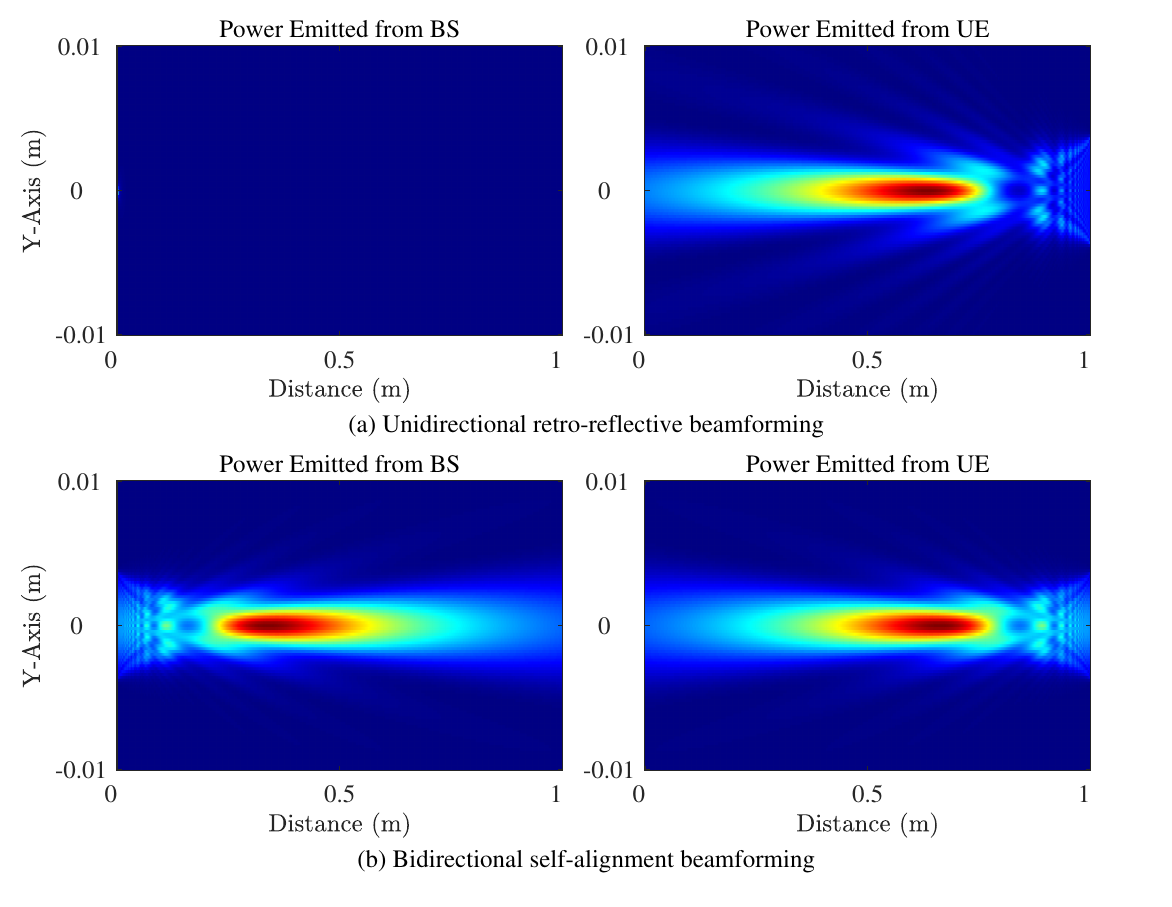}
\caption{Comparison of spatial propagation and beam focusing performance between unidirectional retro-reflective and bidirectional self-alignment systems.}
\label{fig: BS_UE_comparison_expanded}
\end{figure}

\textcolor{blue}{We compare the spatial propagation characteristics of beam patterns between unidirectional retro-reflective system, where only the UE is conjugated, and  bidirectional self-alignment systems, as shown in Fig. \ref{fig: BS_UE_comparison_expanded}. The radiation behavior from both the BS and UE is illustrated. Compared with unidirectional conjugate systems, the proposed bidirectional self-alignment system exhibits significantly enhanced beamfocusing characteristics for signals emitted from both the BS and the UE. As shown in Fig. \ref{fig: BS_UE_comparison_expanded}(a), in unidirectional conjugate beamforming, the signal transmitted from the BS undergoes substantial attenuation during propagation, allowing only a small portion of the power to reach the terminal. Although conjugate reflection at the UE improves the power reflected back to the BS, the resulting beam pattern remains relatively divergent. In contrast, the proposed bidirectional self-alignment beamforming approach, illustrated in Fig. \ref{fig: BS_UE_comparison_expanded}(b), demonstrates superior beam concentration properties in both transmission directions. It significantly reduces spatial divergence during propagation, enabling more focused and efficient energy transmission.}

\begin{figure}[h]\color{blue}
\centering
\includegraphics[width=0.8\linewidth, height=0.2\textheight]{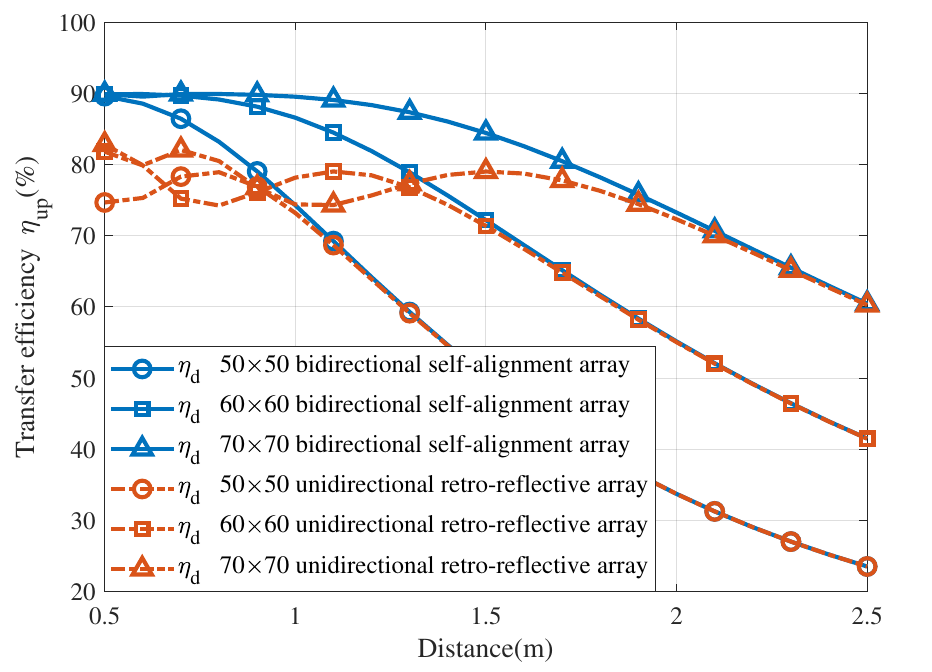}
\caption{Comparison of transmission efficiency performance between unidirectional retro-reflective and bidirectional self-alignment systems.}
\label{fig: Efficiency_Distance_vs_RDA}
\end{figure}

\textcolor{blue}{Fig. \ref{fig: Efficiency_Distance_vs_RDA} presents a comparison of uplink transfer efficiency for arrays of different dimensions. It can be observed that, within the certain transfer distance supported by each array, the uplink transfer efficiency of the proposed bidirectional conjugate system consistently outperforms that of the unidirectional conjugate system. When the transfer distance exceeds the $d_{max}$ of the array, the efficiency of both systems converges. This occurs because the system operates at low power levels where power gain is insufficient to compensate for propagation loss, rendering the power cycle mechanism ineffective. Notably, in the unidirectional retro-reflective system, larger arrays exhibit lower transfer efficiency compared to smaller ones at specific transfer distances. For instance, at $d$ = 0.8 m, the 60 × 60 array demonstrates lower efficiency than the 50 × 50 array. This phenomenon can be attributed to the beam characteristics of larger arrays, although they achieve narrower beam width and higher peak gain, the spatial distribution of } 
\textcolor{blue}{amplitude across the UE array becomes less uniform. This non-uniformity reduces the coherent combining gain during retro-reflection toward the BS, thereby degrading the overall power transfer efficiency. The received power at the UE is proportional to  $(\sum_{m=1}^{M} \sum_{n=1}^{N} \left |E_{_{n} } \right | )^{2} $.}

 \begin{equation*}\color{blue}
     (\sum_{m=1}^{M} \sum_{n=1}^{N} \left |E_{_{n,m} }   \right | )^{2}\le MN\sum_{m=1}^{M} \sum_{n=1}^{N} \left | E_{n,m}  \right | ^{2}.
 \end{equation*}
 
 \textcolor{blue}{According to the Cauchy–Schwarz inequality, the square of the absolute sum of the signal matrix elements is constrained by the sum of the squares of the unit signal energies and the size of the array, as shown in the above equation. Thus, the uniformity of the amplitude distribution significantly affects the received power. The 50 × 50 array, owing to its better amplitude uniformity, achieves higher received power at certain distances compared to the 60 × 60 configuration. However, this phenomenon does not persist when compared to arrays with larger dimensions. }

\section{CONCLUSION}

\textcolor{blue}{In conventional terahertz systems, beam misalignment becomes a critical challenge due to the use of narrow beams and dynamic environmental conditions, which often necessitate frequent and resource-intensive beam training procedures. In this paper, we propose self-alignment mechanism that autonomously establish and maintain beam alignment when the UE moved within the FoV, mitigating the issues of beam misalignment. By incorporating antenna propagation theory and a bidirectional conjugate beamforming structure, we develop the power cycle model to analyze the power distribution and transfer efficiency during the establishment of the self-alignment process. Simulation results demonstrate that under the power cycle model, the signal phases coherently superimpose when they arrive at the BS or UE, forming highly focused communication beams. Furthermore, with the retro-reflected design, the system supports adaptive mobility within a 90{°} range over a 2.2 m transfer distance, enabling free self-alignment within the FoV while achieving a charging power exceeding 9.98 mW and a channel capacity of 127 Gbit/s.}

\textcolor{blue}{The proposed system also can be applied in areas such as smart manufacturing and IoT devices. For example, in high-precision automated production lines, it can provide continuously aligned communication links and stable power supply to robots and industrial sensors, supporting real-time control and sustained operation. Similarly, in smart home or office environments, it can meet the demands of high-speed, low-latency connectivity for mobile devices, including VR headsets, smart appliances, and environmental sensors. Further improvements are warranted in several aspects, including multi-user self-alignment mechanisms, expansion of the dynamic field of view, and practical system implementation.}


\small
\bibliography{mybib}

\begin{thebibliography}{10}
\providecommand{\url}[1]{#1}
\csname url@samestyle\endcsname
\providecommand{\newblock}{\relax}
\providecommand{\bibinfo}[2]{#2}
\providecommand{\BIBentrySTDinterwordspacing}{\spaceskip=0pt\relax}
\providecommand{\BIBentryALTinterwordstretchfactor}{4}
\providecommand{\BIBentryALTinterwordspacing}{\spaceskip=\fontdimen2\font plus
\BIBentryALTinterwordstretchfactor\fontdimen3\font minus \fontdimen4\font\relax}
\providecommand{\BIBforeignlanguage}[2]{{%
\expandafter\ifx\csname l@#1\endcsname\relax
\typeout{** WARNING: IEEEtran.bst: No hyphenation pattern has been}%
\typeout{** loaded for the language `#1'. Using the pattern for}%
\typeout{** the default language instead.}%
\else
\language=\csname l@#1\endcsname
\fi
#2}}
\providecommand{\BIBdecl}{\relax}
\BIBdecl

\bibitem{r1}
M.~Mohammadi, L.-N. Tran, Z.~Mobini, H.~Q. Ngo, and M.~Matthaiou, ``Cell-free massive {MIMO}-assisted {SWIPT} for {IoT} networks,'' \emph{IEEE Transactions on Wireless Communications}, 2025.

\bibitem{r2}
T.~D. Hua, M.~Mohammadi, H.~Q. Ngo, and M.~Matthaiou, ``Cell-free massive {MIMO} {SWIPT} with beyond diagonal reconfigurable intelligent surfaces,'' \emph{IEEE Transactions on Communications}, 2025.

\bibitem{r3}
S.~Thomas, J.~S. Virdi, A.~Babakhani, and I.~P. Roberts, ``A survey on advancements in thz technology for 6{G}: Systems, circuits, antennas, and experiments,'' \emph{IEEE Open Journal of the Communications Society}, 2025.

\bibitem{r4}
R.~Zhang, W.~Wu, X.~Chen, Z.~Gao, and Y.~Cai, ``Terahertz integrated sensing and communication-empowered {UAV}s in 6{G}: {A} transceiver design perspective,'' \emph{IEEE Vehicular Technology Magazine}, 2025.

\bibitem{r5}
C.~Zhu, X.~Xie, C.~Ding, Y.~Zhou, X.~Gao, and J.~An, ``Terahertz empowered vehicular fog computing: opportunities, feasibility, and enhancements,'' \emph{IEEE Wireless Communications}, vol.~31, no.~4, pp. 315--323, 2024.

\bibitem{r6}
S.~Abadal, C.~Han, V.~Petrov, L.~Galluccio, I.~F. Akyildiz, and J.~M. Jornet, ``Electromagnetic nanonetworks beyond 6{G}: {F}rom wearable and implantable networks to on-chip and quantum communication,'' \emph{IEEE Journal on Selected Areas in Communications}, vol.~42, no.~8, pp. 2122--2142, 2024.

\bibitem{r7}
K.~Humadi, I.~Trigui, W.-P. Zhu, and W.~Ajib, ``Terahertz user-centric clustering in the presence of beam misalignment,'' \emph{IEEE Transactions on Vehicular Technology}, vol.~73, no.~8, pp. 12\,115--12\,120, 2024.

\bibitem{r8}
A.~Saeed, O.~Gurbuz, W.~Gerstacker, and I.~Roppelt, ``Impact of beam misalignment fading on terahertz band drone communications,'' \emph{IEEE Wireless Communications Letters}, 2025.

\bibitem{r9}
P.-H. Chang and T.-D. Chiueh, ``Hybrid beamforming for wideband terahertz massive {MIMO} communications with low-resolution phase shifters and true-time-delay,'' \emph{IEEE Transactions on Wireless Communications}, vol.~23, no.~7, pp. 8000--8012, 2024.

\bibitem{r11}
Y.~Gao, Q.~Wu, G.~Zhang, W.~Chen, D.~W.~K. Ng, and M.~Di~Renzo, ``Beamforming optimization for active intelligent reflecting surface-aided {SWIPT},'' \emph{IEEE Transactions on Wireless Communications}, vol.~22, no.~1, pp. 362--378, 2022.

\bibitem{r28}
E.~Shi, J.~Zhang, H.~Du, B.~Ai, C.~Yuen, D.~Niyato, K.~B. Letaief, and X.~Shen, ``{RIS}-aided cell-free massive {MIM}o systems for 6{G}: Fundamentals, system design, and applications,'' \emph{Proceedings of the IEEE}, vol. 112, no.~4, pp. 331--364, 2024.

\bibitem{r46}
C.~Jeong, J.~Park, and H.~Yu, ``Random access in millimeter-wave beamforming cellular networks: issues and approaches,'' \emph{IEEE Communications Magazine}, vol.~53, no.~1, pp. 180--185, 2015.

\bibitem{r47}
F.~J. Martin-Vega, M.~C. Aguayo-Torres, G.~Gomez, J.~T. Entrambasaguas, and T.~Q. Duong, ``Key technologies, modeling approaches, and challenges for millimeter-wave vehicular communications,'' \emph{IEEE Communications Magazine}, vol.~56, no.~10, pp. 28--35, 2018.

\bibitem{r10}
W.~Chen, L.~Li, Z.~Chen, Y.~Liu, B.~Ning, and T.~Q.~S. Quek, ``{ISAC}-enabled beam alignment for terahertz networks: {S}cheme design and coverage analysis,'' \emph{IEEE Transactions on Vehicular Technology}, vol.~73, no.~12, pp. 19\,019--19\,033, 2024.

\bibitem{r12}
J.~Wang, W.~Tang, S.~Jin, C.-K. Wen, X.~Li, and X.~Hou, ``Hierarchical codebook-based beam training for {RIS}-assisted mmwave communication systems,'' \emph{IEEE Transactions on Communications}, vol.~71, no.~6, pp. 3650--3662, 2023.

\bibitem{r13}
C.~Wu, C.~You, Y.~Liu, L.~Chen, and S.~Shi, ``Two-stage hierarchical beam training for near-field communications,'' \emph{IEEE Transactions on Vehicular Technology}, vol.~73, no.~2, pp. 2032--2044, 2024.

\bibitem{r14}
P.~Wang, J.~Fang, W.~Zhang, and H.~Li, ``Fast beam training and alignment for {IRS}-assisted millimeter wave/terahertz systems,'' \emph{IEEE Transactions on Wireless Communications}, vol.~21, no.~4, pp. 2710--2724, 2022.

\bibitem{r15}
W.~Mei and R.~Zhang, ``Intelligent reflecting surface for multi-path beam routing with active/passive beam splitting and combining,'' \emph{IEEE Communications Letters}, vol.~26, no.~5, pp. 1165--1169, 2022.

\bibitem{r48}
W.~{M}ei and R.~Zhang, ``Distributed beam training for intelligent reflecting surface enabled multi-hop routing,'' \emph{IEEE Wireless Communications Letters}, vol.~10, no.~11, pp. 2489--2493, 2021.

\bibitem{r19}
Q.~Liu, M.~Xiong, M.~Liu, Q.~Jiang, W.~Fang, and Y.~Bai, ``Charging a smartphone over the air: The resonant beam charging method,'' \emph{IEEE Internet of Things Journal}, vol.~9, no.~15, pp. 13\,876--13\,885, 2022.

\bibitem{r20}
M.~Liu, S.~Xia, M.~Xiong, M.~Xu, W.~Fang, and Q.~Liu, ``Integrated communication and positioning with resonant beam,'' \emph{IEEE Transactions on Wireless Communications}, vol.~21, no.~11, pp. 9186--9199, 2022.

\bibitem{r21}
S.~Xia, Q.~Jiang, W.~Fang, Q.~Liu, S.~Zhou, M.~Liu, and M.~Xiong, ``Millimeter-wave resonant beam {SWIPT},'' \emph{IEEE Internet of Things Journal}, vol.~11, no.~24, pp. 40\,464--40\,477, 2024.

\bibitem{r22}
M.~Xiong, Q.~Liu, X.~Wang, S.~Zhou, B.~Zhou, and Z.~Bu, ``Mobile optical communications using second harmonic of intra-cavity laser,'' \emph{IEEE Transactions on Wireless Communications}, vol.~21, no.~5, pp. 3222--3231, 2021.

\bibitem{r23}
W.~Fang, W.~Chen, Q.~Wu, K.~Wang, S.~Zhang, Q.~Liu, and J.~Li, ``Reconfigurable intelligent surface assisted free space optical information and power transfer,'' \emph{IEEE Internet of Things Journal}, vol.~11, no.~18, pp. 30\,260--30\,277, 2024.

\bibitem{r24}
D.~Dardari, M.~Lotti, N.~Decarli, and G.~Pasolini, ``Grant-free random access with backscattering self-conjugating metasurfaces,'' \emph{IEEE Transactions on Cognitive Communications and Networking}, 2024.

\bibitem{r25}
{Dardari, Davide and Lotti, Marina and Decarli, Nicol{\`o} and Pasolini, Gianni}, ``Establishing {MIMO} communications automatically using self-conjugating metasurfaces,'' in \emph{ICC 2023-IEEE International Conference on Communications}.\hskip 1em plus 0.5em minus 0.4em\relax IEEE, 2023, pp. 1286--1292.

\bibitem{r26}
R.~Miyamoto and T.~Itoh, ``Retrodirective arrays for wireless communications,'' \emph{IEEE Microwave Magazine}, vol.~3, no.~1, pp. 71--79, 2002.

\bibitem{r27}
S.-C. Yen and T.-H. Chu, ``A retro-directive antenna array with phase conjugation circuit using subharmonically injection-locked self-oscillating mixers,'' \emph{IEEE Transactions on Antennas and Propagation}, vol.~52, no.~1, pp. 154--164, 2004.

\bibitem{r29}
H.-B. Jung and J.-H. Lee, ``Theoretical and experimental investigation of {N}-bit reconfigurable retrodirective metasurface,'' \emph{Journal of Electromagnetic Engineering and Science}, vol.~24, no.~1, pp. 51--56, 2024.

\bibitem{r30}
D.~M. Pozar, \emph{Microwave engineering: {t}heory and techniques}.\hskip 1em plus 0.5em minus 0.4em\relax John wiley \& sons, 2021.

\bibitem{r31}
W.~Jiang, Q.~Zhou, J.~He, M.~A. Habibi, S.~Melnyk, M.~El-Absi, B.~Han, M.~Di~Renzo, H.~D. Schotten, F.-L. Luo \emph{et~al.}, ``Terahertz communications and sensing for 6{G} and beyond: A comprehensive review,'' \emph{IEEE Communications Surveys \& Tutorials}, vol.~26, no.~4, pp. 2326--2381, 2024.

\bibitem{r32}
H.~T. Friis, ``A note on a simple transmission formula,'' \emph{Proceedings of the IRE}, vol.~34, no.~5, pp. 254--256, 1946.

\bibitem{r33}
H.~Shi, C.~Yang, and M.~Peng, ``Comprehensive link-level simulator for terahertz {mimo} integrated sensing and communication systems with tdd framework,'' \emph{IEEE Transactions on Vehicular Technology}, vol.~73, no.~11, pp. 16\,932--16\,947, 2024.

\bibitem{r34}
W.~Tang, J.~Y. Dai, M.~Z. Chen, K.-K. Wong, X.~Li, X.~Zhao, S.~Jin, Q.~Cheng, and T.~J. Cui, ``{MIMO} transmission through reconfigurable intelligent surface: System design, analysis, and implementation,'' \emph{IEEE journal on selected areas in communications}, vol.~38, no.~11, pp. 2683--2699, 2020.

\bibitem{r35}
B.~Ning, Z.~Tian, W.~Mei, Z.~Chen, C.~Han, S.~Li, J.~Yuan, and R.~Zhang, ``Beamforming technologies for ultra-massive {MIMO} in terahertz communications,'' \emph{IEEE Open Journal of the Communications Society}, vol.~4, pp. 614--658, 2023.

\bibitem{r36}
S.~Mizojiri, K.~Takagi, K.~Shimamura, S.~Yokota, M.~Fukunari, Y.~Tatematsu, and T.~Saito, ``Ga{N} schottky barrier diode for sub-terahertz rectenna,'' in \emph{2019 IEEE Wireless Power Transfer Conference (WPTC)}, 2019, pp. 36--39.

\bibitem{r37}
W.~Tang, X.~Li, J.~Y. Dai, S.~Jin, Y.~Zeng, Q.~Cheng, and T.~J. Cui, ``Wireless communications with programmable metasurface: Transceiver design and experimental results,'' \emph{China Communications}, vol.~16, no.~5, pp. 46--61, 2019.

\bibitem{r38}
W.~Tang, J.~Y. Dai, M.~Chen, X.~Li, Q.~Cheng, S.~Jin, K.-K. Wong, and T.~J. Cui, ``Programmable metasurface-based {RF} chain-free {8PSK} wireless transmitter,'' \emph{Electronics letters}, vol.~55, no.~7, pp. 417--420, 2019.

\bibitem{r39}
Z.~Zhang, L.~Dai, X.~Chen, C.~Liu, F.~Yang, R.~Schober, and H.~V. Poor, ``Active {RIS} vs. {P}assive {RIS}: Which will prevail in 6{G}?'' \emph{IEEE Transactions on Communications}, vol.~71, no.~3, pp. 1707--1725, 2023.

\bibitem{r40}
X.~Li, W.~Chen, P.~Zhou, Y.~Wang, F.~Huang, S.~Li, J.~Chen, and Z.~Feng, ``A 250–310 {GH}z power amplifier with 15-d{B} peak gain in 130-nm {S}i{G}e {B}i{CMOS} process for terahertz wireless system,'' \emph{IEEE Transactions on Terahertz Science and Technology}, vol.~12, no.~1, pp. 1--12, 2022.

\bibitem{r41}
C.~A. Balanis, \emph{Antenna {T}heory: {A}nalysis and Design}.\hskip 1em plus 0.5em minus 0.4em\relax Hoboken, NJ, USA: Wiley, 2016.

\bibitem{r42}
Q.~Li, ``Millimeter wave propagation loss in the atmosphere and the effect on 5{G} communication,'' Master's thesis, Xidian University, Xian, China, May 2017, in Chinese.

\bibitem{r45}
\BIBentryALTinterwordspacing
M.~Fujishima and S.~Amakawa, \emph{Design of Terahertz CMOS Integrated Circuits for High-Speed Wireless Communication}.\hskip 1em plus 0.5em minus 0.4em\relax The Institution of Engineering and Technology, 2019. [Online]. Available: \url{https://digital-library.theiet.org/doi/abs/10.1049/PBCS035E}
\BIBentrySTDinterwordspacing

\bibitem{r44}
K.~Zhi, C.~Pan, H.~Ren, K.~K. Chai, and M.~Elkashlan, ``Active {RIS} versus passive {RIS}: Which is superior with the same power budget?'' \emph{IEEE Communications Letters}, vol.~26, no.~5, pp. 1150--1154, 2022.

\bibitem{r43}
M.~Najafi, V.~Jamali, R.~Schober, and H.~V. Poor, ``Physics-based modeling and scalable optimization of large {I}ntelligent {R}eflecting {S}urfaces,'' \emph{IEEE Transactions on Communications}, vol.~69, no.~4, pp. 2673--2691, 2021.

\end{thebibliography}

\end{document}